\documentclass[preprintnumbers,nofootinbib]{revtex4}
\pdfoutput=1

\usepackage{graphicx}
\usepackage{amsmath}
 \usepackage{amssymb}

\allowdisplaybreaks

\newcommand{\be}{\begin{equation}}
\newcommand{\ee}{\end{equation}}
\newcommand{\bea}{\begin{eqnarray}}
\newcommand{\eea}{\end{eqnarray}}

\newcommand{\Z}{\mathbb{Z}}
\newcommand{\N}{\mathbb{N}}

\newcommand{\bmu}{\mbox{\boldmath$\mu$}}

\begin{document}
\title{Phase structure of two-dimensional QED at zero
  temperature with flavor-dependent chemical
  potentials and the role of multidimensional theta functions}
\date{\today}
\author{Robert Lohmayer}
\email{robert.lohmayer@fiu.edu}
\author{Rajamani Narayanan}
\email{rajamani.narayanan@fiu.edu}
\affiliation{ 
Department of Physics, Florida International University,
Miami, FL 33199, USA.}

\begin{abstract}
We consider QED on a two-dimensional Euclidean torus with $f$ flavors
of massless
fermions and flavor-dependent chemical potentials. The dependence of
the partition function on the chemical potentials is reduced to a
$(2f-2)$-dimensional theta function. At zero temperature, the system can exist in an infinite number
of
phases characterized by certain values of traceless
particle numbers  and separated by first-order phase transitions.
Furthermore, there exist many points in the $(f-1)$-dimensional space of
traceless chemical potentials where two or three phases can coexist for $f=3$ and
 two, three, four or six phases can coexist for $f=4$. We conjecture that the maximal number of coexisting phases grows exponentially with increasing $f$. 
\end{abstract}

\preprint{\today}

\maketitle

\section{Introduction and Summary}
QED in two dimensions is a useful toy model to gain an understanding
of the theory at finite temperature and chemical potential~\cite{Sachs:1993zx,Sachs:1995dm,Sachs:1991en}. In
particular,
the physics at zero temperature is interesting since one can study a
system that can exist in several phases.
The theory at zero temperature is governed by two degrees of freedom
often referred to as the toron variables in a Hodge decomposition of
the U(1) gauge field on a $l\times \beta$ torus where $l$ is the
circumference of the spatial circle and $\beta$ is the inverse
temperature. Integrating over the toron fields projects on to a state
with net zero charge~\cite{Gross:1980br} and therefore there is no dependence on a flavor-independent chemical potential~\cite{Narayanan:2012du}. The dependence
on the isospin chemical potential for the two flavor case was studied
in~\cite{Narayanan:2012qf}
and we extend this result to the case of $f$ flavors in this paper. 
After integrating out the toron variables, the dependence on the
$(f-1)$ traceless\footnote{linear combinations that are invariant under uniform (flavor-independent) shifts} chemical potential variables and the dimensionless
temperature $\tau=\frac{l}{\beta}$ can be written in the
form of a $(2f-2)$-dimensional theta function (see~\cite{multitheta}
for an overview on multidimensional theta functions). 
The $(2f-2)$ dimensional theta function has a non-trivial Riemann
matrix and this is a consequence of the same gauge field (toron
variables, in particular) that couples to all flavors.
The
resulting phase structure is quite intricate since it involves
minimization of a quasi-periodic function over a set of integers.
We will explicitly show: 
\begin{enumerate}
\item {\sl Three flavors}: The two-dimensional plane defined
  by the two traceless chemical potentials is filled by hexagonal
  cells (c.f.~Fig.~\ref{fig:f3-phases} in this paper) with the system having a specific value of the two traceless
  particle numbers  in each cell and neighboring cells being separated by first-order phase transitions at zero temperature. The vertices of the hexagon are
  shared
by three cells and therefore two or three different phases can coexist at
zero temperature.
\item {\sl Four flavors}: The three-dimensional space defined by the three
  traceless chemical potentials is filled by two types of
  cells (c.f.~Fig.~\ref{fig:f4-cells} in this paper). One
  of them can be viewed as a cube with the edges cut off. We then stack many of these cells such that they
  join at the square faces. The remaining space is filled by
  the second type of cell. All edges of either one of the cells are
  shared by three cells
but we have two types of vertices -- one type shared by four cells and
another shared by six cells. At zero temperature, each cell can be identified by a unique
value for the three different traceless particle numbers  and neighboring cells are separated by first-order phase transitions. Therefore, two, three, four or six phases can coexist at
zero temperature.
\end{enumerate}
One can use the multidimensional theta function to study the phase
structure
when $f>4$ but visualization of the cell structure becomes difficult.
Nevertheless, it is possible to provide examples of the coexistence of many phases.
We conjecture that the maximal number of coexisting phases is given by
$f \choose{ \lfloor f/2 \rfloor} $, 
increasing exponentially for large $f$.

The organization of the paper is as follows. We derive the dependence
of the partition function on the $(f-1)$ traceless chemical potentials
and
the dimensionless temperature $\tau$ in section~\ref{pf}. 
We briefly show the connection to the two flavor case discussed
in~\cite{Narayanan:2012qf}
and focus in detail on the three and four flavor cases in
section~\ref{results}.
We then conclude the paper with a discussion of some examples when $f
> 4$.

\section{The partition function}\label{pf}

Consider $f$-flavored massless QED on a finite torus with spatial
length $l$ and dimensionless temperature $\tau$. All flavors have
the same gauge coupling $\frac{e}{l}$ where $e$ is dimensionless.
Let 
\be
\bmu^t = \begin{pmatrix}
\mu_1 & \mu_2 & \cdots  & \mu_f\cr
\end{pmatrix}
\ee
be the flavor-dependent chemical potential vector.
The partition function 
is~\cite{Sachs:1991en,Narayanan:2012qf} 
\be
Z(\bmu,\tau,e) = Z_b(\tau,e)
Z_t(\bmu,\tau),\label{partfun}
\ee
where the bosonic part is given by
\be
Z_b(\tau,e) = \frac{1}{\eta^{2f}(i\tau)}\sideset{}{'}\prod_{k_1,k_2=-\infty}^\infty
\frac{1}{\sqrt{ 
\left(k_2^2 +\frac{1}{\tau^2}k_1^2\right)
\left(k_2^2 +\frac{1}{\tau^2}\left[
    k_1^2+\frac{fe^2}{4\pi^3}\right]\right)}}\label{bosonpf}
\ee
(with $k_1=k_2=0$ excluded from the product and $\eta(i\tau)$ being the
 Dedekind eta function) and the toronic part reads
\begin{align}
Z_t(\bmu,\tau) &=
\int_{-\frac{1}{2}}^{\frac{1}{2}} dh_2
\int_{-\frac{1}{2}}^{\frac{1}{2}} dh_1\, \prod_{i=1}^f g(h_1,h_2,\tau,\mu_i)\,,\cr
g(h_1,h_2,\tau,\mu) &=
\sum_{n,m=-\infty}^\infty
\exp \left[ -\pi\tau \left[\left(n+ h_2 -i \frac{\mu}{\tau}\right)^2
+\left(m + h_2 -i \frac{\mu}{\tau}\right)^2\right] +2\pi i h_1 \left(n -m\right)\right].\label{toronpf}
\end{align}
We will only consider ourselves with the physics at zero temperature
and therefore focus on the toronic part and
perform the integration over the toronic variables, $h_1$ and
$h_2$.
\vskip 1cm

\subsection{Multidimensional theta function}

\begin{statement}

The toronic part of the partition function has a representation in the
form of a $(2f-2)$-dimensional theta function:
\be 
Z_t(\bmu,\tau)=\frac{1}{\sqrt{2\tau f}}
\sum_{{\boldsymbol n}=-\infty}^\infty 
\exp \left[  -\pi\tau \left({\boldsymbol n}^t T^t +\frac{i}{\tau} {\boldsymbol s}^t\right) 
\begin{pmatrix}
\bar\Omega & {\boldsymbol 0} \cr
{\boldsymbol 0} & \bar\Omega\cr
\end{pmatrix}
\left( T {\boldsymbol n} + \frac{i}{\tau}{\boldsymbol s}\right) \right]\label{maineqn}
\ee
where
${\boldsymbol n}$ is a $(2f-2)$-dimensional vector of integers.
The $(2f-2)\times (2f-2)$ transformation matrix $T$ is
\begin{align} 
T &= 
\begin{pmatrix}
1  & 0 & \cdots & 0 & 0\cr
0  & 1 & \cdots & 0 & 0\cr
0  & 0 & \cdots & 0 & 0\cr
0  & 0 & \cdots & 1 & 0\cr
-1  & -1 & \cdots & -1 & f\cr
\end{pmatrix};&
T^{-1} &= 
\begin{pmatrix}
1  & 0 & \cdots & 0 & 0\cr
0  & 1 & \cdots & 0 & 0\cr
0  & 0 & \cdots & 0 & 0\cr
0  & 0 & \cdots & 1 & 0\cr
\frac{1}{f}  & \frac{1}{f} & \cdots & \frac{1}{f} & \frac{1}{f}\cr
\end{pmatrix}.\label{transt}
\intertext{The $(f-1)\times (f-1)$ matrix $\bar\Omega$ is}
\bar\Omega &= 
\begin{pmatrix}
1 - \frac{1}{f} & -\frac{1}{f} & \cdots & -\frac{1}{f} \cr
- \frac{1}{f} & 1-\frac{1}{f} & \cdots & -\frac{1}{f}\cr
\vdots & \vdots & \ddots & \vdots \cr
- \frac{1}{f} & -\frac{1}{f} & \cdots & 1-\frac{1}{f}\cr
\end{pmatrix}\,; 
&\bar\Omega^{-1} &= 
\begin{pmatrix}
2 & 1 & \cdots & 1 \cr
1 & 2 & \cdots & 1\cr
\vdots & \vdots & \ddots & \vdots \cr
1 & 1 & \cdots & 2\cr
\end{pmatrix}\,;\cr
\bar\Omega &= R
\begin{pmatrix}
1 & 0 & \cdots 0 & 0\cr
0 & 1 & \cdots 0 & 0 \cr
\vdots & \vdots & \ddots & \vdots & \vdots \cr
0 & 0 & \cdots 1 & 0\cr
0 & 0 & \cdots 0 & \frac{1}{f}\cr
\end{pmatrix}
R^t\,,
&R_{ij} &=
\begin{cases}
\frac{1}{\sqrt{j(j+1)}} & i \le j < (f-1)\cr
-\frac{j}{\sqrt{j(j+1)}} & i = j+1 \le (f-1)\cr
0 & i > j+1 \le (f-1)\cr
\frac{1}{\sqrt{f-1}} & j=(f-1);\ \ \ \forall i\cr
\end{cases}\,\,.
\label{omega}
\end{align}
 The dependence on the chemical potentials comes from
\be\label{eq:s}
{\boldsymbol s}^t = 
\begin{pmatrix}
\bar \mu_2 &
\bar \mu_3 &
\cdots &
\bar \mu_f &
-\bar \mu_2 &
-\bar \mu_3 &
\cdots &
-\bar \mu_f \cr
\end{pmatrix}
\ee
where we have separated the chemical potentials into a flavor-independent component and $(f-1)$ traceless components using
\be\label{eq:mubar}
\begin{pmatrix}
\bar \mu_1 \cr
\bar \mu_2 \cr
\vdots \cr
\bar \mu_f \cr
\end{pmatrix} =
M\bmu\,,
\qquad\qquad
M = 
\begin{pmatrix}
 1 & 1 & 1 & \cdots & 1\cr
 1 & -1 & 0 & \cdots & 0\cr
 1 & 0 & -1 & \cdots & 0\cr
 \vdots & \vdots & \vdots & \ddots & \vdots\cr
 1 & 0 & 0 & \cdots & -1\cr
\end{pmatrix}.
\ee 
\end{statement}

\begin{proof}

Consider the sum
\be
{\boldsymbol a}^t {\boldsymbol a} = \sum_{i=1}^f a_i a_i.
\ee
Noting that
\be 
N=\begin{pmatrix}
 1 & 1 & 1 & \cdots & 1\cr
 1 & -(f-1) & 1 & \cdots & 1\cr
 1 & 1 & -(f-1) & \cdots & 1\cr
 \vdots & \vdots & \vdots & \ddots & \vdots\cr
 1 & 1 & 1 & \cdots & -(f-1)\cr
\end{pmatrix}\,, \qquad\qquad
NM=f,
\ee
it follows that 
\be
{\boldsymbol a}^t {\boldsymbol a}=
\frac{1}{f} \sum_{i=1}^f b_i \bar a_i\qquad \text{with} \qquad \boldsymbol{\bar a}=M{\boldsymbol a}\,,\quad
\boldsymbol{b}=N\boldsymbol{a}\,.
\ee
Explicitly,
\be 
b_1 = \bar a_1 \qquad \text{and} \qquad
b_i = \bar a_1 -fa_i = f\bar a_i
-\sum_{j=2}^f \bar a_j\,,\qquad
i=2,\ldots, f\,,
\ee 
where we have used the relation
\be
f a_1 = \sum_{i=1}^f \bar a_i\,.\label{aabar}
\ee
Therefore,
\be
{\boldsymbol a}^t {\boldsymbol a}=\frac 1{f^2} \boldsymbol{\bar a}^t N^2 \boldsymbol{\bar a} =
\frac{1}{f} \bar a_1^2  -\frac{1}{f} \sum_{i,j=2}^f \bar a_i \bar a_j
+\sum_{i=2}^f \bar a_i^2\,.
\ee
Setting
\be
\boldsymbol{\bar n} = M{\boldsymbol n}\,,\qquad \boldsymbol{\bar m} = M{\boldsymbol m}\,,\qquad \boldsymbol{\bar \bmu} = M\bmu
\ee
in (\ref{toronpf}) and using the relation (\ref{aabar}) to rewrite $\bar n_1$
and $\bar m_1$,
we obtain
\begin{align}
Z_t(\bmu,\tau) &=
\sum_{n_1,m_1,\{\bar n_i,\bar m_i\}=-\infty}^\infty 
\int_{-\frac{1}{2}}^{\frac{1}{2}} dh_2
\int_{-\frac{1}{2}}^{\frac{1}{2}} dh_1\,
\exp\left[2\pi i h_1 \left(f \left(n_1-m_1\right) -\sum_{i=2}^f \left(\bar n_i - \bar m_i\right)\right)\right]\times
\cr 
&\quad \times
\exp \Biggl[ -\pi\tau\Biggl(
\frac{1}{f} \left\{f n_1 -\sum_{i=2}^f \bar n_i+fh_2
    -i\frac{\bar \mu_1}{\tau}\right\}^2
+ \frac 1f \left\{f m_1  -\sum_{i=2}^f \bar m_i+ fh_2 -i\frac{\bar \mu_1}{\tau}\right\}^2\cr
& \quad\quad\quad\quad\quad\quad\quad
-\frac{1}{f} \sum_{i,j=2}^f \left\{
\left(\bar n_i  -i\frac{\bar \mu_i}{\tau}\right) \left(\bar n_j  -i\frac{\bar \mu_j}{\tau}\right)
+\left(\bar m_i  -i\frac{\bar \mu_i}{\tau}\right) \left(\bar m_j
 -i\frac{\bar \mu_j}{\tau}\right)\right\}\cr
& \quad\quad\quad\quad\quad\quad\quad
+\sum_{i=2}^f \left\{\left(\bar n_i  -i\frac{\bar \mu_i}{\tau}\right)^2
+\left(\bar m_i  -i\frac{\bar \mu_i}{\tau}\right)^2\right\}
\Biggr)\Biggr]
\end{align}
where $n_1$, $m_1$, $\bar n_i$ and $\bar m_i$,
$i=2,\ldots, f$, are the new set of summation variables.
The integral over $h_1$ results in 
\begin{align}
Z_t(\bmu,\tau) =
\sideset{}{'}\sum_{n_1,\{\bar n_i,\bar m_i\}=-\infty}^\infty 
\int_{-\frac{1}{2}}^{\frac{1}{2}} dh_2
\exp \Biggl[ -\pi\tau\Biggl(&
\frac{2}{f} \left\{f n_1 -\sum_{i=2}^f \bar n_i+fh_2 -i\frac{\bar \mu_1}{\tau}\right\}^2\cr
&
-\frac{1}{f} \sum_{i,j=2}^f \left\{
\left(\bar n_i  -i\frac{\bar \mu_i}{\tau}\right) \left(\bar n_j  -i\frac{\bar \mu_j}{\tau}\right)
+\left(\bar m_i  -i\frac{\bar \mu_i}{\tau}\right) \left(\bar m_j
-i\frac{\bar \mu_j}{\tau}\right)\right\}\cr
&
+\sum_{i=2}^f \left\{\left(\bar n_i  -i\frac{\bar \mu_i}{\tau}\right)^2
+\left(\bar m_i  -i\frac{\bar \mu_i}{\tau}\right)^2\right\}
\Biggr)\Biggr]\,,
\end{align}
where the prime denotes that $\sum_{i=2}^f \left(\bar n_i -\bar
  m_i\right)$ be a multiple of $f$.
The integral over $h_2$ along with the sum over $n_1$ reduces to a
complete Gaussian integral and the result is
\begin{align}
Z_t(\bmu,\tau) =\frac{1}{\sqrt{2\tau f}}
\sideset{}{'}\sum_{\{\bar n_i,\bar m_i\}=-\infty}^\infty 
\exp \Biggl[ -\pi\tau\Biggl(&
\sum_{i=2}^f \left\{\left(\bar n_i  -i\frac{\bar \mu_i}{\tau}\right)^2
+\left(\bar m_i  -i\frac{\bar \mu_i}{\tau}\right)^2\right\}\cr
&
-\frac{1}{f} \sum_{i,j=2}^f \left\{
\left(\bar n_i  -i\frac{\bar \mu_i}{\tau}\right) \left(\bar n_j  -i\frac{\bar \mu_j}{\tau}\right)
+\left(\bar m_i  -i\frac{\bar \mu_i}{\tau}\right) \left(\bar m_j
-i\frac{\bar \mu_j}{\tau}\right)\right\}
\Biggr)\Biggr]\label{maineqn1}\,.
\end{align}
The prime in the sum can be removed if we trade $\bar n_f$ for $\bar k$ 
where
\be
\bar n_f = \sum_{i=2}^f \bar m_i -\sum_{i=2}^{f-1} \bar n_i + \bar k f.\label{maineqn2}
\ee 
We change $\bar m_i\to-\bar m_i$ and
define the $(2f-2)$-dimensional vector 
\be
{\boldsymbol n}^t = 
\begin{pmatrix}
\bar m_2 &
\bar m_3 &
\cdots &
\bar m_f &
\bar n_2 &
\bar n_3 &
\cdots &
\bar n_{f-1} &
\bar k \cr
\end{pmatrix}.
\ee 
Then statement (\ref{maineqn}) follows from (\ref{maineqn1}).
\end{proof}

\subsection{Particle number }

We define particle numbers $N_i$ corresponding to the chemical potentials $\mu_i$ as
\begin{align}
  N_i(\bmu,\tau) = \frac{\tau}{4\pi}\frac{\partial}{\partial \mu_i} \ln Z_t(\bmu,\tau)\,.
\end{align}
Analogously to Eq.~\eqref{eq:mubar}, we set
\begin{align}
   \bar N_k(\bmu,\tau) = N_1(\bmu,\tau)-N_k(\bmu,\tau) \qquad\text{for}\ 2\leq k \leq f\,.
\end{align}
In the infinite-$\tau$ limit, the infinite sums in Eq.~\eqref{maineqn} are dominated by ${\boldsymbol n}={\boldsymbol 0}$ which results in 
\begin{align}
   \bar N_k(\bmu,\infty) = \bar \mu_k \qquad\text{for}\ 2\leq k \leq f\,.\label{numdeninf}
\end{align}
Since the partition function is independent of $\bar \mu_1$, $\bar N_1(\bmu,\tau)=\sum_{i=1}^f N_i(\bmu,\tau)=0$ for all $\tau$.

\subsection{Zero-temperature limit}\label{sec:lowT}

In order to study the physics at zero temperature ($\tau\to 0$) we
set
\be
\Omega = T^t \begin{pmatrix}
\bar\Omega & {\boldsymbol 0} \cr
{\boldsymbol 0} & \bar\Omega\cr
\end{pmatrix} T;\qquad
\Gamma = \frac{1}{\tau}
T^{-1}\,.
\ee
Then we can rewrite (\ref{maineqn}) using
the Poisson summation formula as
\be
Z_t(\bmu,\tau) 
=
\frac{1}{\sqrt{2\tau f}\tau^{f-1}}
\sum_{{\boldsymbol k}=-\infty}^\infty 
\exp \left[ -\frac{\pi}{\tau}\left( {\boldsymbol k}^t
\Omega^{-1} {\boldsymbol k} -2 {\boldsymbol k}^t T^{-1}
{\boldsymbol s}\right)
\right]
\label{zffinal}
\ee
with
\be
\frac{1}{\Omega} = 
\begin{pmatrix}
2 & 1 & \cdots & 1& 1 & 0 & 0 & \cdots & 0 & 1\cr
1 & 2 & \cdots  & 1& 1 & 0 & 0 & \cdots & 0 & 1\cr
\vdots & \vdots & \ddots & \vdots & \vdots & \vdots & \vdots & \ddots
& \vdots & \vdots\cr
1 & 1 & \cdots & 2 & 1 & 0 & 0 & \cdots 0 & 0 & 1\cr 
1 & 1 & \cdots & 1 & 2 & 0 & 0 & \cdots 0 & 0 & 1\cr 
0 & 0 & \cdots & 0 & 0 & 2 & 1 & \cdots & 1 &1 \cr
0 & 0 & \cdots & 0 & 0 & 1 & 2 & \cdots  &1 &1 \cr
\vdots & \vdots & \ddots & \vdots & \vdots & \vdots & \vdots & \ddots
& \vdots & \vdots \cr
0 & 0 & \cdots & 0 &0 & 1 & 1 & \cdots & 2 & 1 \cr
1 & 1 & \cdots & 1 &1 & 1 & 1 & \cdots & 1 & 2-\frac{2}{f} \cr
\end{pmatrix}\label{oinverse}\,,
\ee
where the block in the upper left corner has dimensions $(f-1)\times(f-1)$ and the second block on the diagonal has dimensions $(f-2)\times(f-2)$\,.

For fixed $\bar \mu_k$, the partition function in the zero-temperature
limit is determined by minimizing the term $ {\boldsymbol k}^t
\Omega^{-1} {\boldsymbol k} -2 {\boldsymbol k}^t T^{-1}
{\boldsymbol s}$ in the exponent in Eq.~\eqref{zffinal}. 
Assuming in general that the minimum is $M$-fold degenerate, let
$S=\{{\boldsymbol k}^{(i)}\}_{i=1,\ldots,M}$, ${\boldsymbol
  k}^{(i)}\in \Z^{2f-2}$, label these $M$ minima. Then
\begin{align}
\bar N_j(\bmu,0) &=\frac1{2M} \sum_{i=1}^M \left(\sum_{l=1}^{f-1} k^{(i)}_l - \sum_{l=f}^{2f-3} k^{(i)}_l+k^{(i)}_{j-1} -k^{(i)}_{f+j-2}\right)\,,\qquad 2\leq j\leq f-1\,,\\
\bar N_f(\bmu,0) &=\frac1{2M} \sum_{i=1}^M \left(\sum_{l=1}^{f-1} k^{(i)}_l - \sum_{l=f}^{2f-3} k^{(i)}_l+k^{(i)}_{f-1}\right)\,.
\end{align}
If the minimum is non-degenerate (or if all ${\boldsymbol k}^{(i)}$ individually result in the same $\bar N_j(\bmu,0) $'s), the particle numbers  $\bar N_j(\bmu,0) $ assume integer or half-integer values at zero temperature. 
Since ${\boldsymbol k}\in \Z^{2f-2}$ and we only have $(f-1)$ $\bar
  N_j(\bmu,0) $ (with $\bar N_1(\bmu,\tau)=0$ for all $\tau$), there
  are in general many possibilities to obtain 
identical particle numbers from different $\boldsymbol k$'s. 
The zero-temperature  phase boundaries in the $(f-1)$-dimensional
space of traceless chemical potentials $\bar \mu_{2,\ldots,f}$ are
determined by those $\boldsymbol{\bar \mu}$'s leading to degenerate minima
with different $\boldsymbol{\bar  N}$'s. As we will see later, phases
with different particle numbers  will be separated by first-order phase transitions. 

One can numerically determine the phase boundaries as follows: Having
chosen one set for the traceless chemical potentials, one finds the 
traceless particle numbers  at zero temperature (by numerically searching
for the minimum) at several points in the traceless chemical potential
space
close to the initial one. We label the initial choice of chemical potentials by
the number of different values one obtains for the traceless particle numbers
 in its small neighborhood and this enables us to trace the
phase boundaries. Whereas this method works in general, it is possible
to perform certain orthogonal changes of variables in the space of
traceless
chemical potentials and obtain expressions equivalent to
(\ref{zffinal})
that are easier to deal with when tracing the phase boundaries. Such
equivalent
expressions for the case of $f=3$ and $f=4$ are provided in Appendix~\ref{sec:Zalt}.

Consider the system at high temperature with a certain choice of
traceless chemical potentials which results in  average values for the
traceless particle numbers
 equal to the choice as per (\ref{numdeninf}). The system
will show typical thermal fluctuations as one cools the system but the
thermal fluctuations will only die down and produce a uniform
distribution of traceless particle numbers  if the initial choice of
traceless chemical potentials did not lie at a point in the phase boundary. Tuning
the
traceless chemical potentials to lie at a point in the phase boundary
will
result in a system at zero temperature with several co-existing
phases. In other words, the system will exhibit spatial
inhomogeneities.
We will demonstrate this for $f=2,3,4$ in section~\ref{results}.

\subsection{Quasi-periodicity}

Consider the change of variables
\be
{\boldsymbol s}' = {\boldsymbol s} + T \Omega^{-1} {\boldsymbol m}
\ee
with ${\boldsymbol m} \in \Z^{2f-2}$.
Since ${\boldsymbol s}$ is of the special form \eqref{eq:s},
there is a restriction on ${\boldsymbol m}$.
From Eq.~\eqref{oinverse}, we find that ${\boldsymbol m}$ has to satisfy
\bea
m_{f-1+k}&=&m_{f-1}-m_k\qquad\qquad 1\leq k \leq f-2\,,\cr
m_{2f-2}&=&-\frac f2 m_{f-1} \ \in \Z\,.\ \label{meqn}
\eea
This corresponds to
\begin{align}\label{eq:shift-mu}
{\bar \mu_{k+1}}'= \bar \mu_{k+1} + m_{k}-\frac f2 m_{f-1}+\sum_{i=1}^{f-1} m_i\qquad\qquad 1\leq k \leq f-1\,,
\end{align}
and
\begin{align}
Z_t(\bmu',\tau) = Z_t(\bmu,\tau) e^{\frac \pi \tau \left({\boldsymbol m}^t \Omega^{-1} {\boldsymbol m}+2{\boldsymbol m}^t T^{-1}{\boldsymbol s}\right)}.
\end{align}
The particle numbers  under this shift are related by
\begin{align}
\bar N_{k+1}(\bmu',\tau) = \bar N_{k+1}(\bmu,\tau) + m_k -\frac{f}{2} m_{f-1} +\sum_{i=1}^{f-1} m_i,
\end{align}
which is the same as the shift in $\bar\mu$ as defined in (\ref{eq:shift-mu}).

\section{Results}\label{results}

\subsection{\boldmath Phase structure for $f=2$}

\begin{figure}[ht]
\centerline{
\includegraphics[width=130mm]{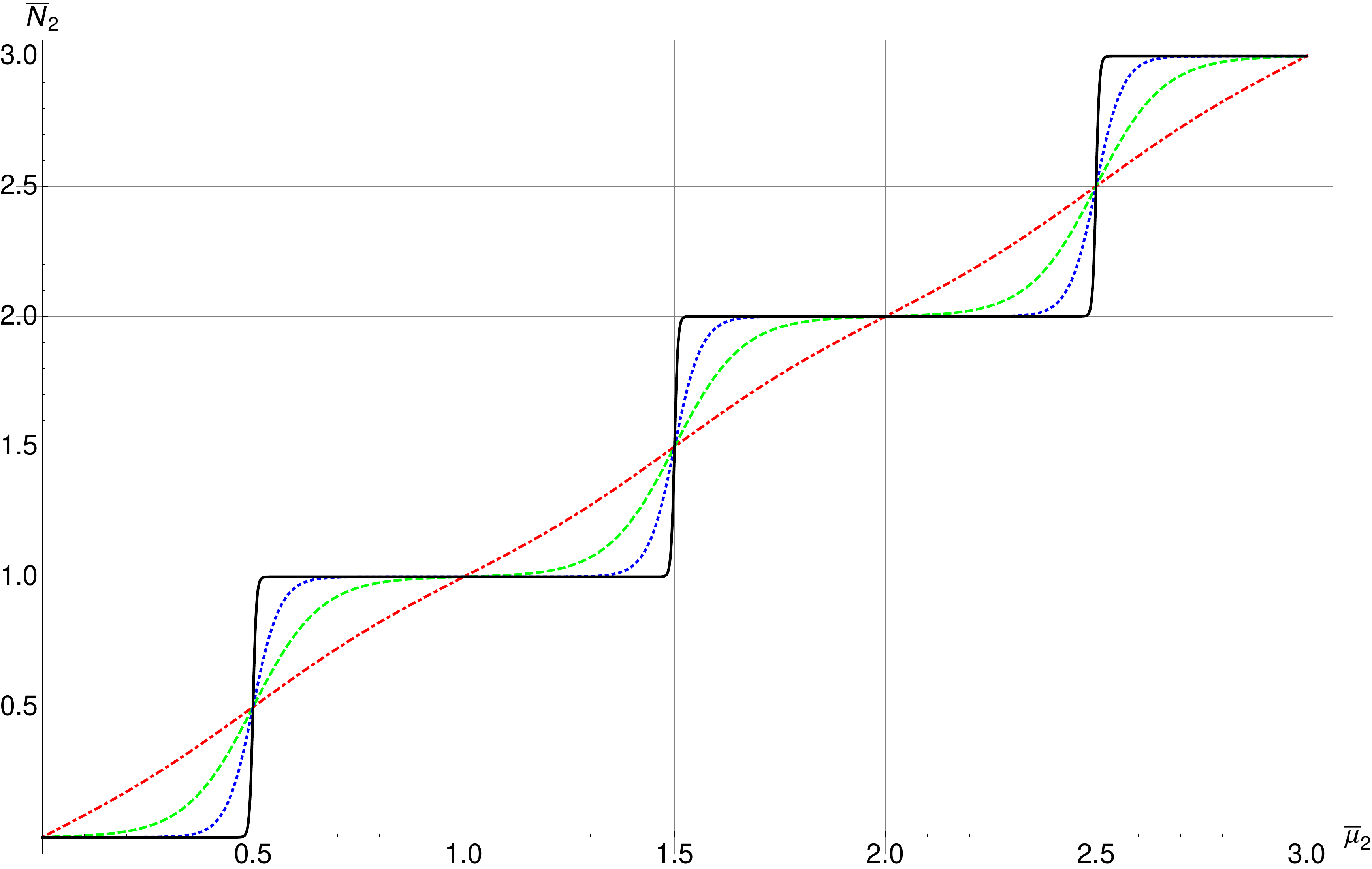}
}\caption{
For $f=2$, plot of $\bar N_2$ as a function of $\bar \mu_2$ for $\tau=1.5$ (red, dotdashed), $\tau=0.5$ (green, dashed), $\tau=0.2$ (blue, dotted), and $\tau=0.025$ (black, solid). } \label{fig:f2}
\end{figure}

We reproduce the results in \cite{Narayanan:2012qf} in this
subsection.
The condition on integer shifts in (\ref{meqn}) reduces to $m_2=-m_1$
and the shift in chemical potential is given by $\bar\mu_2' =
\bar\mu_2 + m_1$. 
From Eq.~\eqref{zffinal} for $f=2$, we obtain
\begin{align}
\bar N_2 = \frac{\sum_{k=-\infty}^\infty k e^{-\frac{\pi}{\tau}\left(k- \bar\mu_2\right)^2}}{\sum_{k=-\infty}^\infty e^{-\frac{\pi}{\tau}\left(k-\bar\mu_2\right)^2}}\,,
\end{align}
and this is plotted in Fig.~\ref{fig:f2}. The quasi-periodicity under $\bar\mu_2' =
\bar\mu_2 + m_1$ is evident.
For small $\tau$, the dominating term in the infinite sum is obtained when $k$ assumes the integer value closest to $\bar\mu_2$. Therefore, $\bar N_2(\bar \mu_2)$ approaches a step function in the zero-temperature limit (see Fig.~\ref{fig:f2}).
Taking into account the first sub-leading term, we obtain (for non-integer $\bar\mu_2$)
\begin{align}
\bar N_2 = \left\lfloor \bar\mu_2 \right\rfloor + \frac 12 \left[1+\tanh\left(\frac \pi \tau \left[\bar \mu_2-\left\lfloor \bar\mu_2 \right\rfloor-\frac12\right]\right)\right]+\ldots
\end{align}
At zero temperature, first-order phase transitions occur at all half-integer values of $\bar \mu_2$, separating phases which are characterized by different (integer) values of $\bar N_2$.

\begin{figure}[htb]
\centering
\includegraphics[width=0.3 \textwidth]{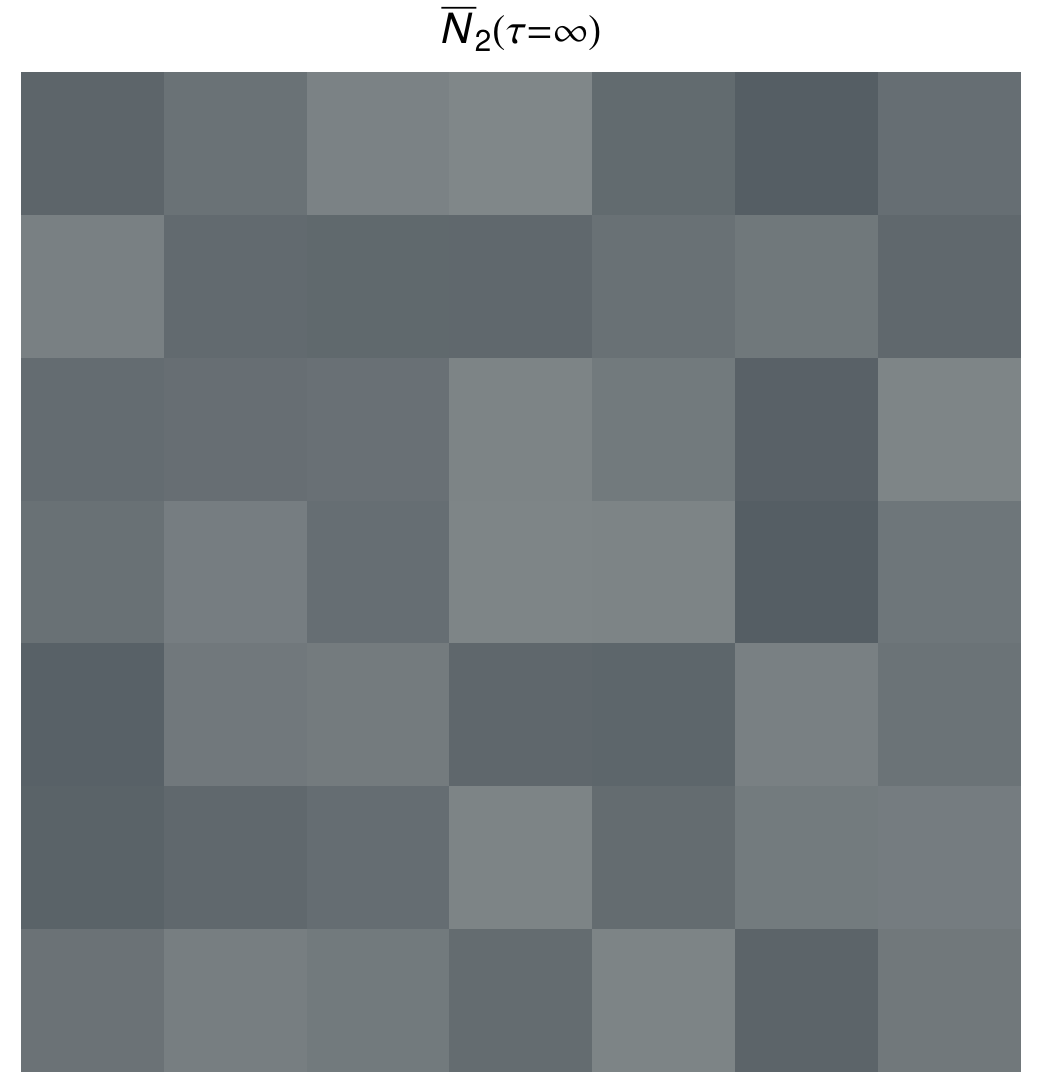}\qquad
\includegraphics[width=0.3 \textwidth]{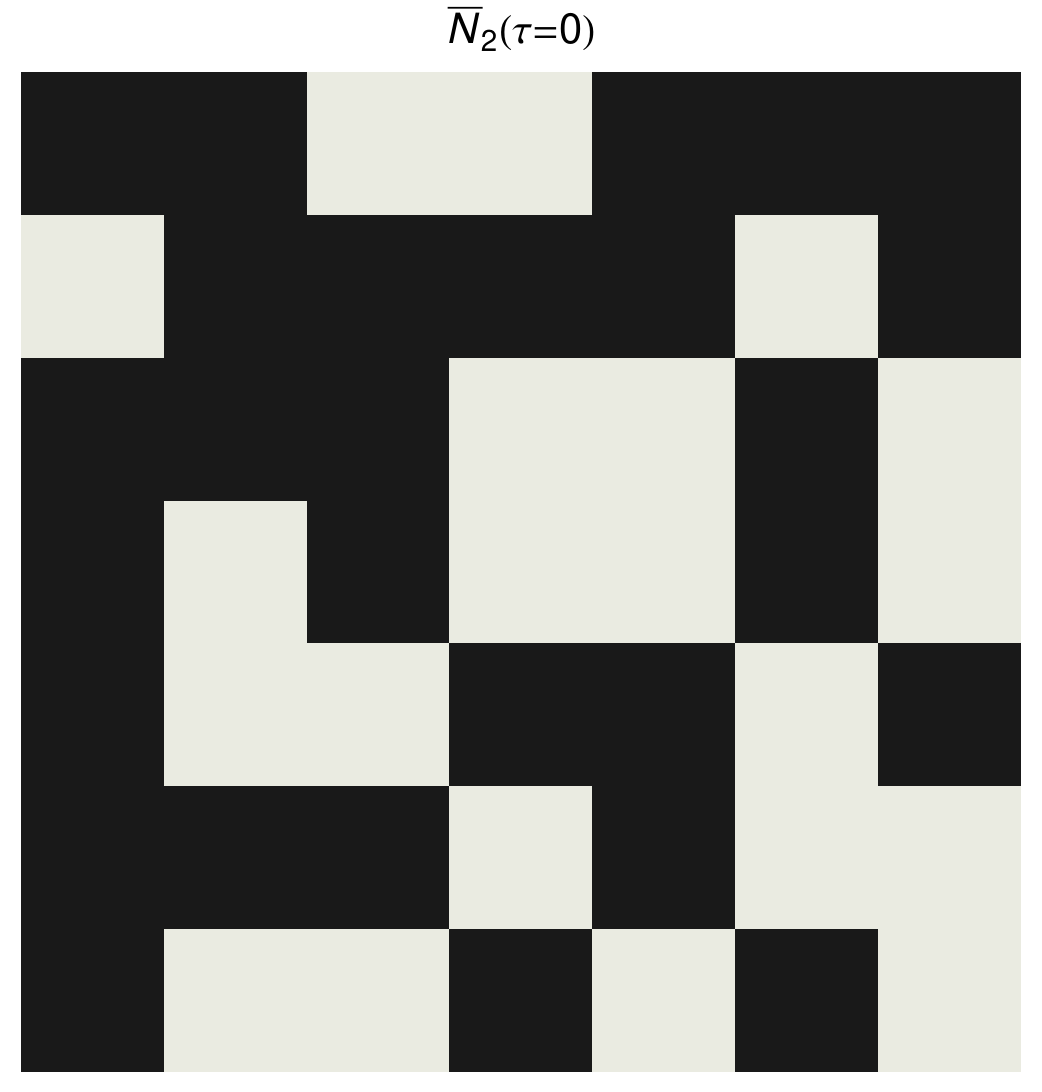}\qquad
\includegraphics[width=0.044\textwidth]{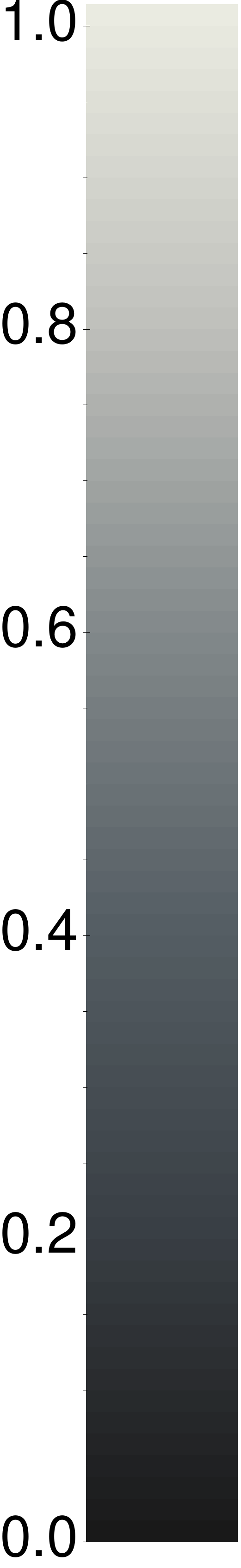}
\caption{For $f=2$, a spacetime grid with small  fluctuations around $\bar N_2=1/2$ at large $\tau$ (left panel) results in two coexisting phases (characterized by $\bar N_2=0$ and $\bar N_2=1$) at zero temperature (right panel).}
\label{fig:f2-grid-inhom}
\end{figure}

\begin{figure}[htb]
\centering
\includegraphics[width=0.3 \textwidth]{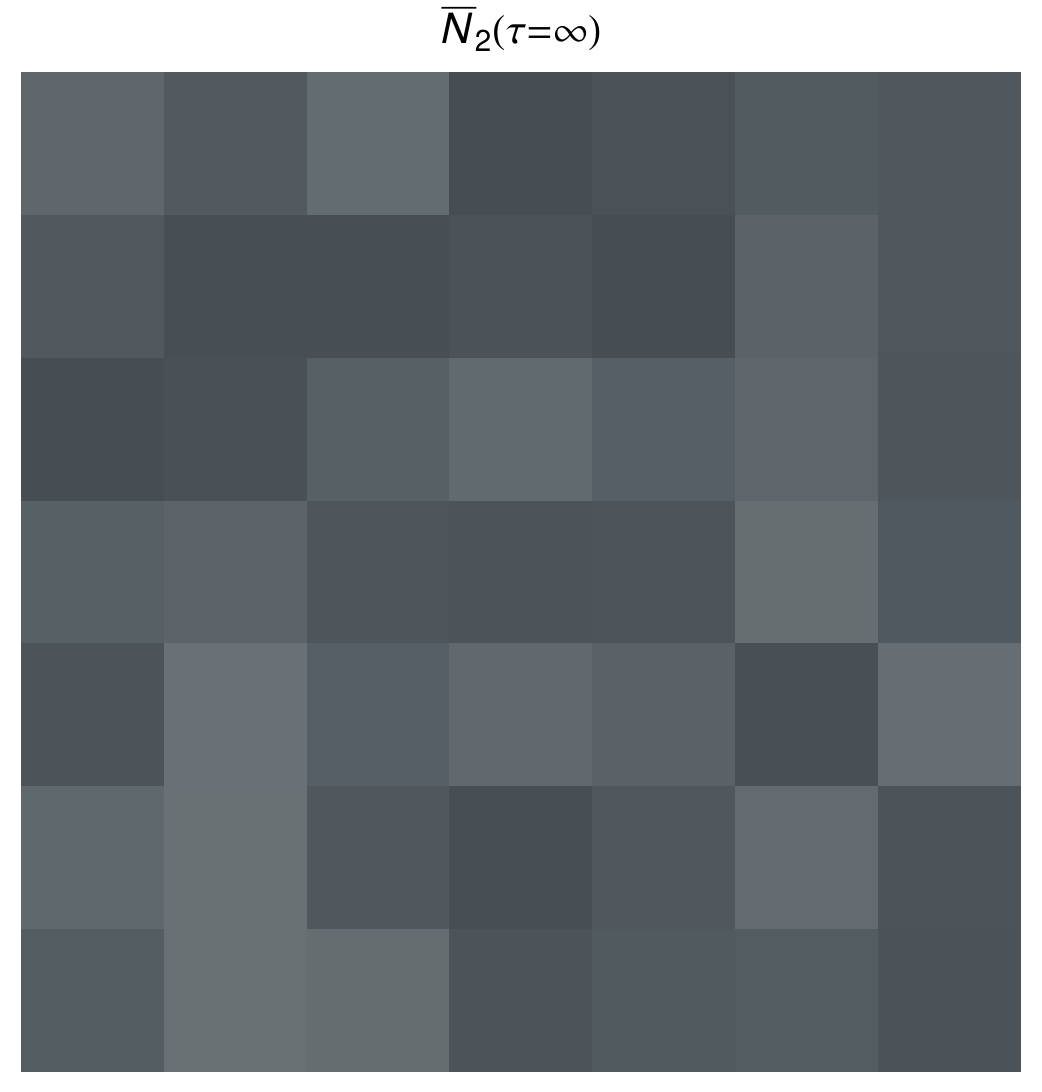}\qquad
\includegraphics[width=0.3 \textwidth]{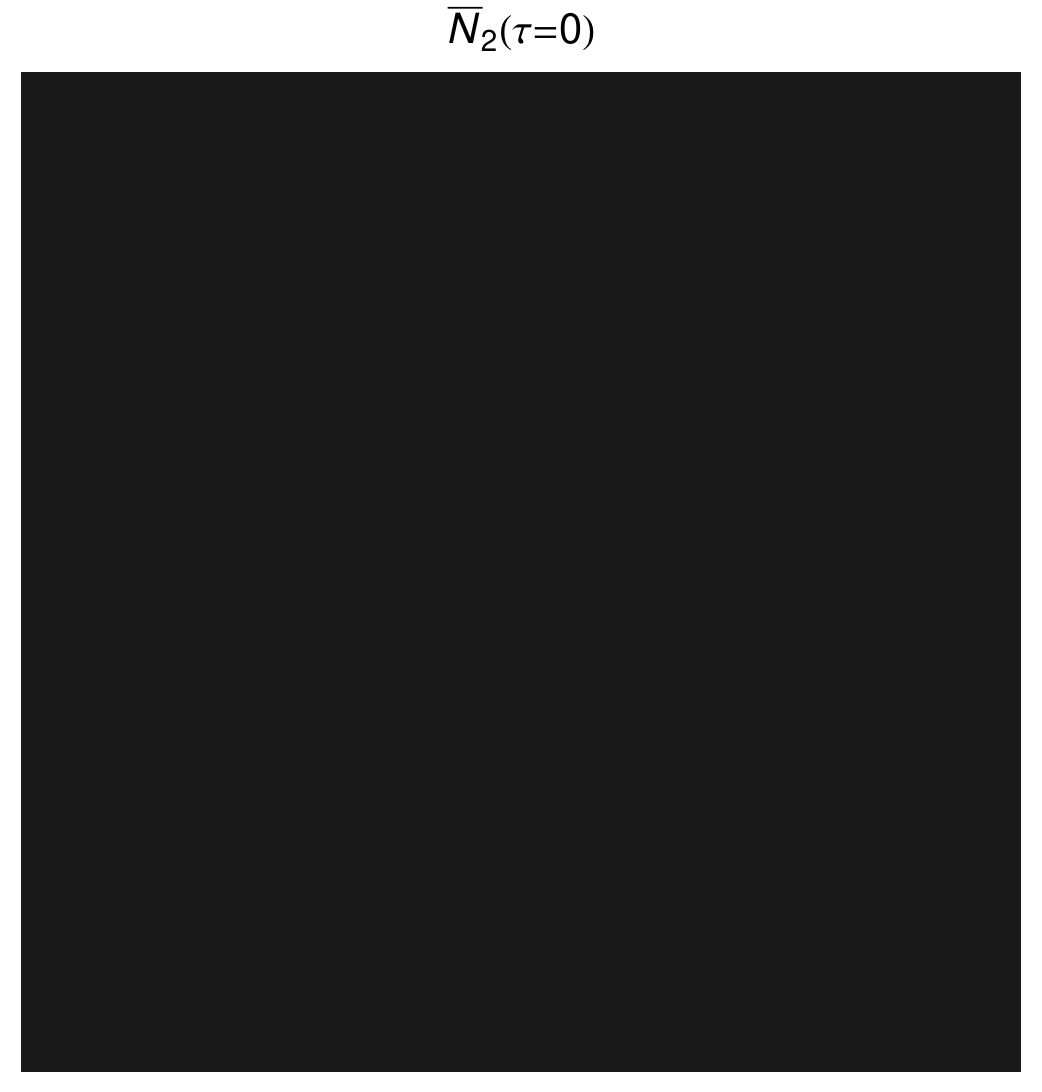}\qquad
\includegraphics[width=0.044\textwidth]{plots/f2-grid-legend}
\caption{For $f=2$, a spacetime grid with random fluctuations around $\bar N_2=2/5$ at large $\tau$ (left panel) results in a uniform particle number  ($\bar N_2=0$) at $\tau=0$ (right panel).}
\label{fig:f2-grid-hom}
\end{figure}

If a system at high temperature is described in the path-integral
formalism by fluctuations (as a function of the two Euclidean
spacetime coordinates) of $\bar N_2$ around a half-integer value, the
corresponding system at zero temperature will have two coexisting
phases (fluctuations are amplified when $\tau$ is decreased). On the
other hand, away from the phase boundaries, the system will become
uniform at zero temperature (fluctuations are damped when $\tau$ is
decreased). Fig.~\ref{fig:f2-grid-inhom} shows spatial inhomogeneities
develop in a 
system with $\bar\mu_2$ chosen at the phase boundary as it is cooled
and Fig.~\ref{fig:f2-grid-hom} shows thermal fluctuations dying down in a
system
with $\bar\mu_2$ chosen away from the phase boundary.  
The square grid with many cells can either be thought of as an
Euclidean spacetime grid or a sampling of several identical systems
(in terms of the choice of $\bar\mu_2$ and $\tau$).

\subsection{\boldmath Phase structure for $f=3$}

\begin{figure}[htb]
\centering
\includegraphics[width=0.42\textwidth]{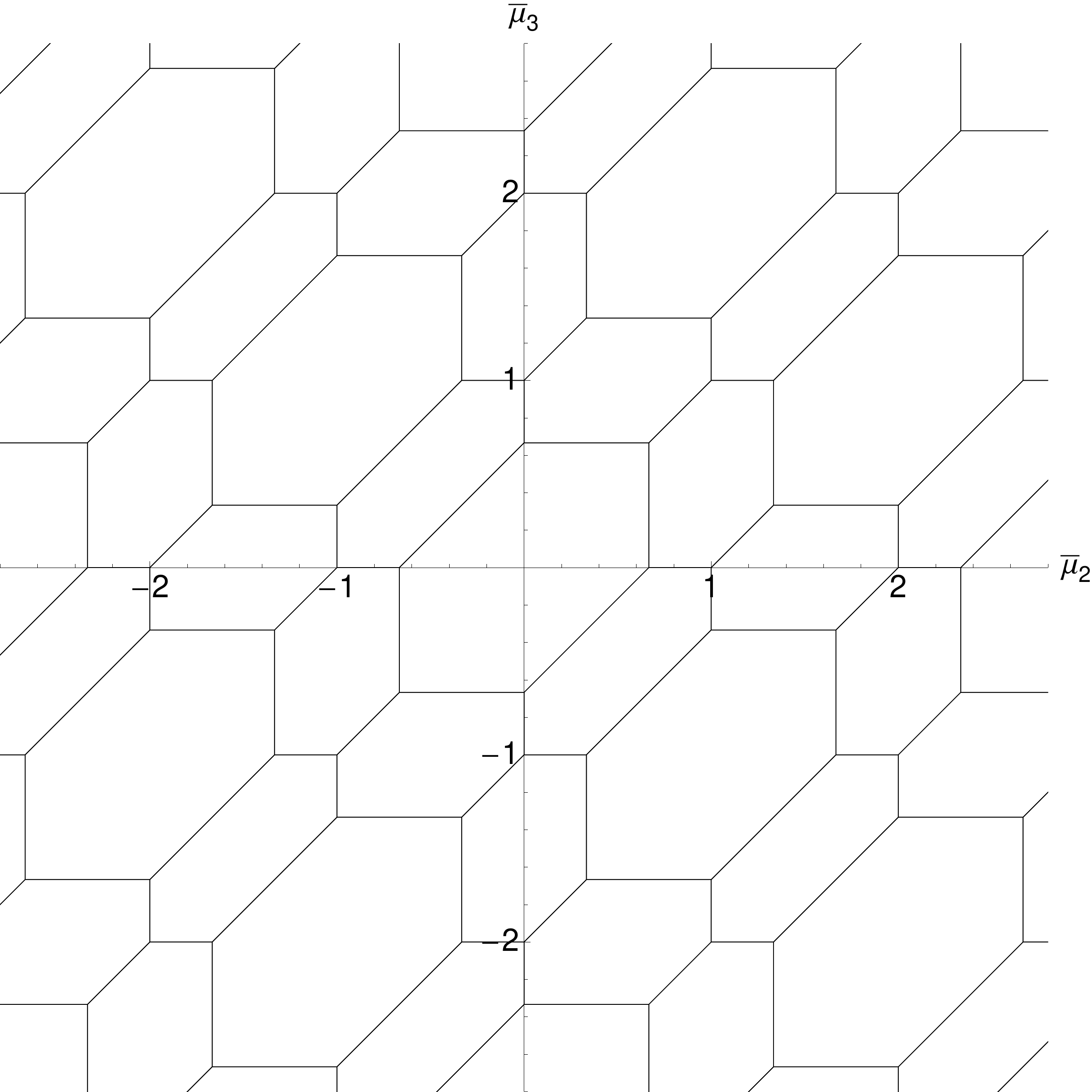}\qquad\qquad
\includegraphics[width=0.42\textwidth]{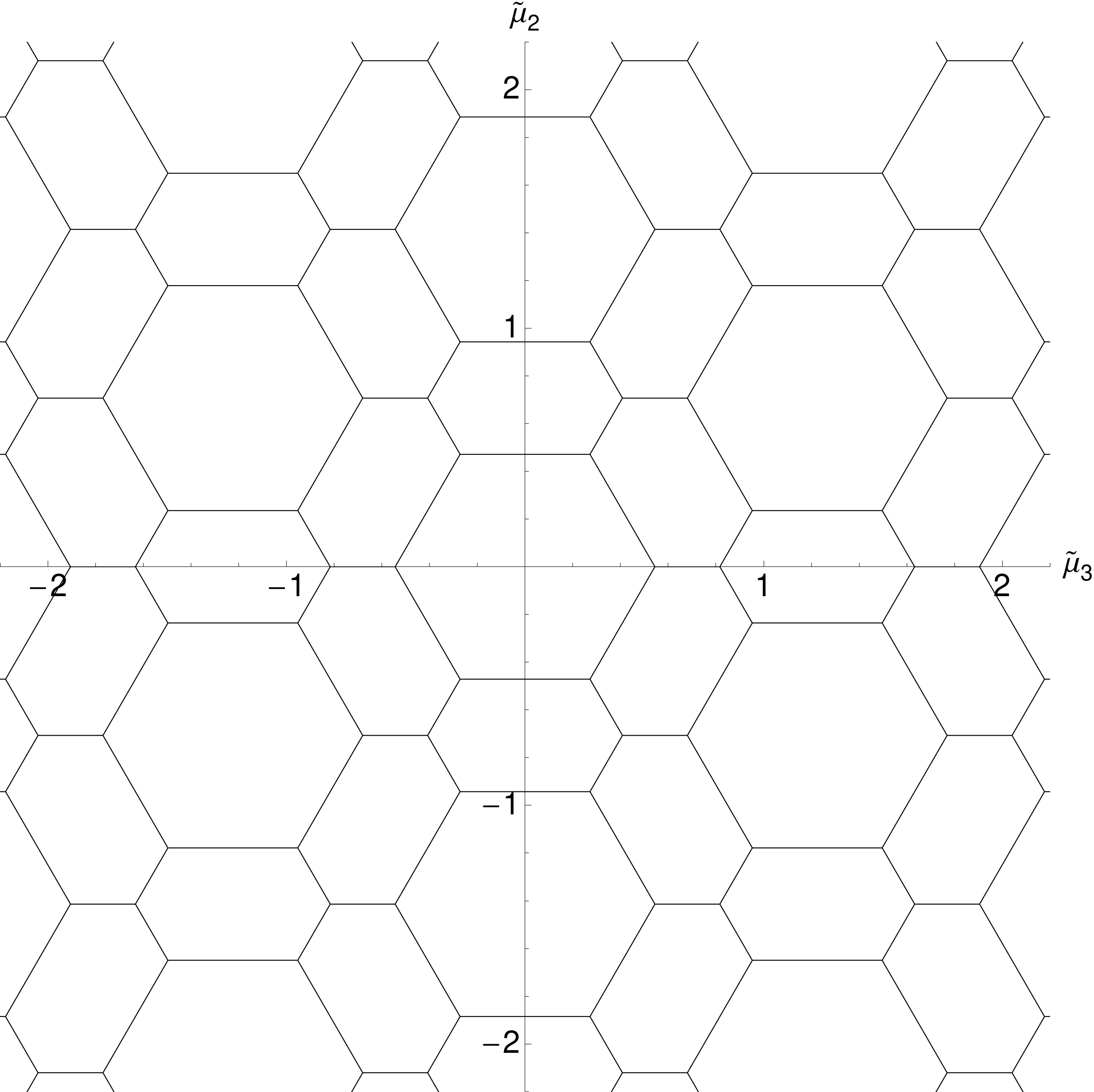}
\caption{Phase boundaries at zero temperature for $f=3$ in the $\bar \mu$ plane (left) and the $\tilde \mu$ plane (right).}
\label{fig:f3-phases}
\end{figure}

We determine the phase boundaries, separating cells with different $(\bar N_2,\bar N_3)$ as described in Sec.~\ref{sec:lowT}. 
As explained in Sec.~\ref{sec:lowT} it is also instructive to use a different coordinate system for the chemical potentials, obtained from $(\mu_1,\mu_2,\mu_3)$ by an orthonormal transformation:  
\begin{align}
\begin{pmatrix}
\tilde \mu_1 \\
\tilde \mu_2 \\
\tilde \mu_3 
\end{pmatrix} 
=
\begin{pmatrix}
\frac 1{\sqrt{3}} & \frac 1{\sqrt{3}} & \frac 1{\sqrt{3}} \\
\frac 1{\sqrt{2}} & -\frac {1}{\sqrt{2}} & 0 \\
\frac 1{\sqrt{6}} & \frac 1{\sqrt{6}} & -\frac 2{\sqrt{6}} \\
\end{pmatrix}
\begin{pmatrix}
\mu_1 \\
\mu_2 \\
\mu_3 
\end{pmatrix} \,,
\end{align} 
i.e., $\tilde \mu_2 = \bar \mu_2 /\sqrt{2}$ and $\tilde \mu_3=(-\bar
\mu_2+2 \bar \mu_3)/\sqrt{6}$. We denote the corresponding particle 
numbers by $\tilde N_2$ and $\tilde N_3$.  An alternative representation of the partition function, which simplifies the determination of vertices in terms of the coordinates $\tilde \mu_i$, is given in appendix \ref{sec:Zalt}. 
In these coordinates, the phase structure is symmetric under rotations
by $\pi/3$ and composed of two types of hexagonal cells, a central
regular hexagon is surrounded by six smaller non-regular hexagons,
which are identical up to rotations.
Figure~\ref{fig:f3-phases} shows the phase boundaries at zero
temperature in both coordinate systems.

The condition on the integers ${\boldsymbol m}$ as given in (\ref{meqn}) reduce to $m_3 =
m_2-m_1$ and $m_4 = -\frac{3}{2} m_2$. Therefore, we require $m_2$ to
be even and write it as $2l_2$. 
From Eq.~\eqref{eq:shift-mu} we see that the boundaries in the $(\bar \mu_2,\bar \mu_3)$ plane are periodic under shifts
\begin{align}\label{eq:f3-shift}
\begin{pmatrix}
\bar \mu'_2 \\
\bar \mu'_3 
\end{pmatrix} =
\begin{pmatrix}
\bar \mu_2 \\
\bar \mu_3 
\end{pmatrix}
+
m_1
\begin{pmatrix}
2 \\
1 
\end{pmatrix} 
-
l_2
\begin{pmatrix}
1 \\
-1 
\end{pmatrix} 
\qquad\qquad m_1,\,l_{2}\in\Z\,.
\end{align}
The shift symmetry \eqref{eq:f3-shift} is obvious in Fig.~\ref{fig:f3-phases}. 

\begin{figure}[htb]
\centering
\includegraphics[width=0.3\textwidth]{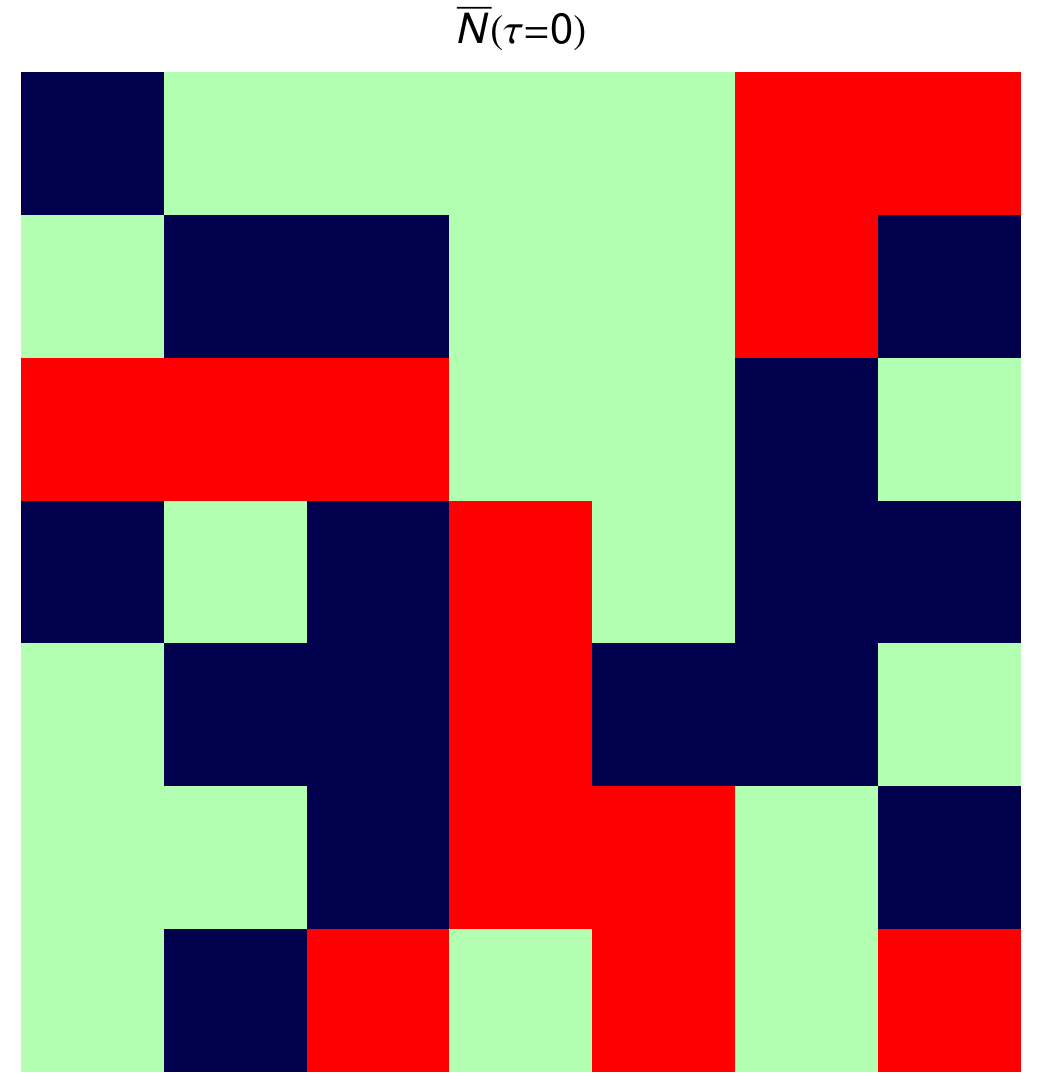}\qquad
\includegraphics[width=0.3\textwidth]{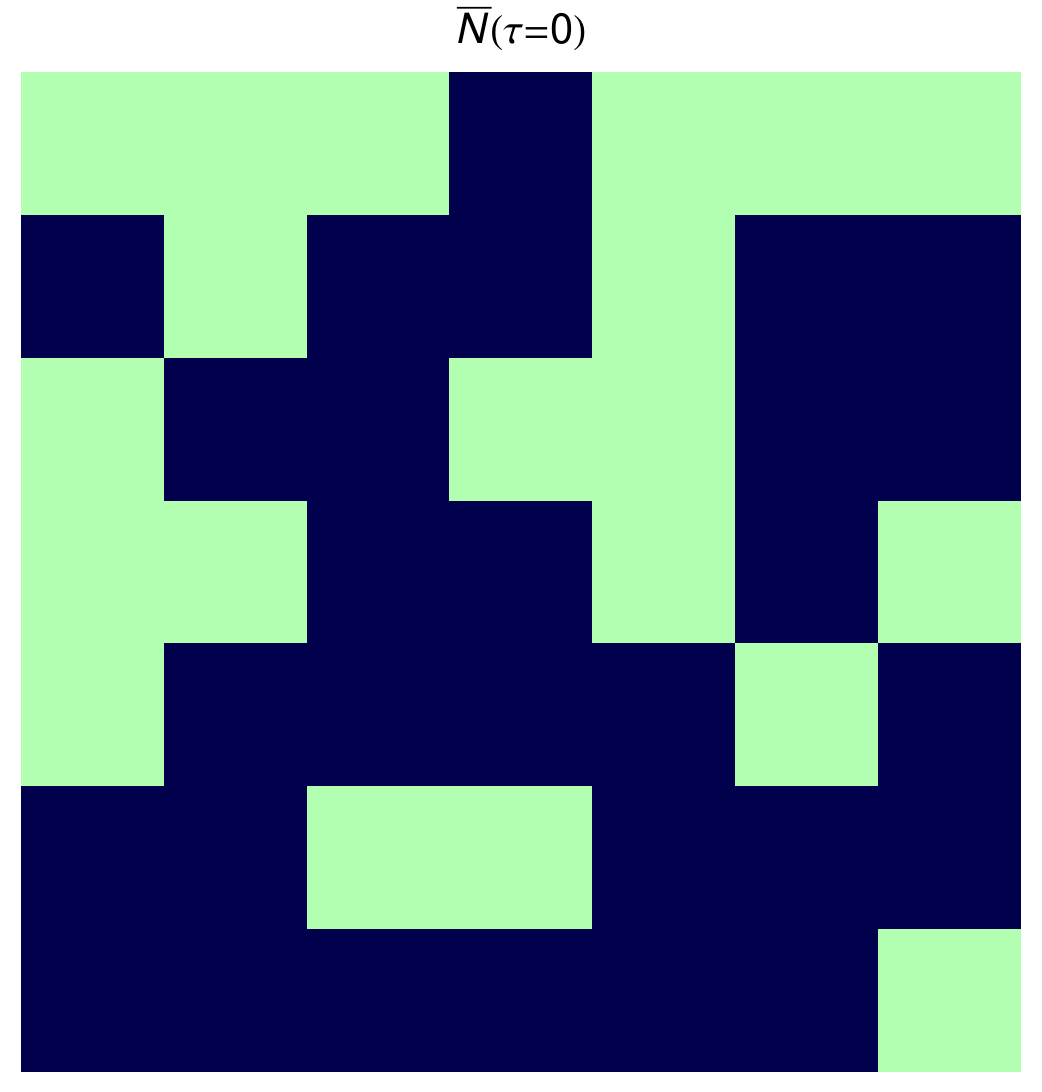}\qquad
\includegraphics[width=0.3\textwidth]{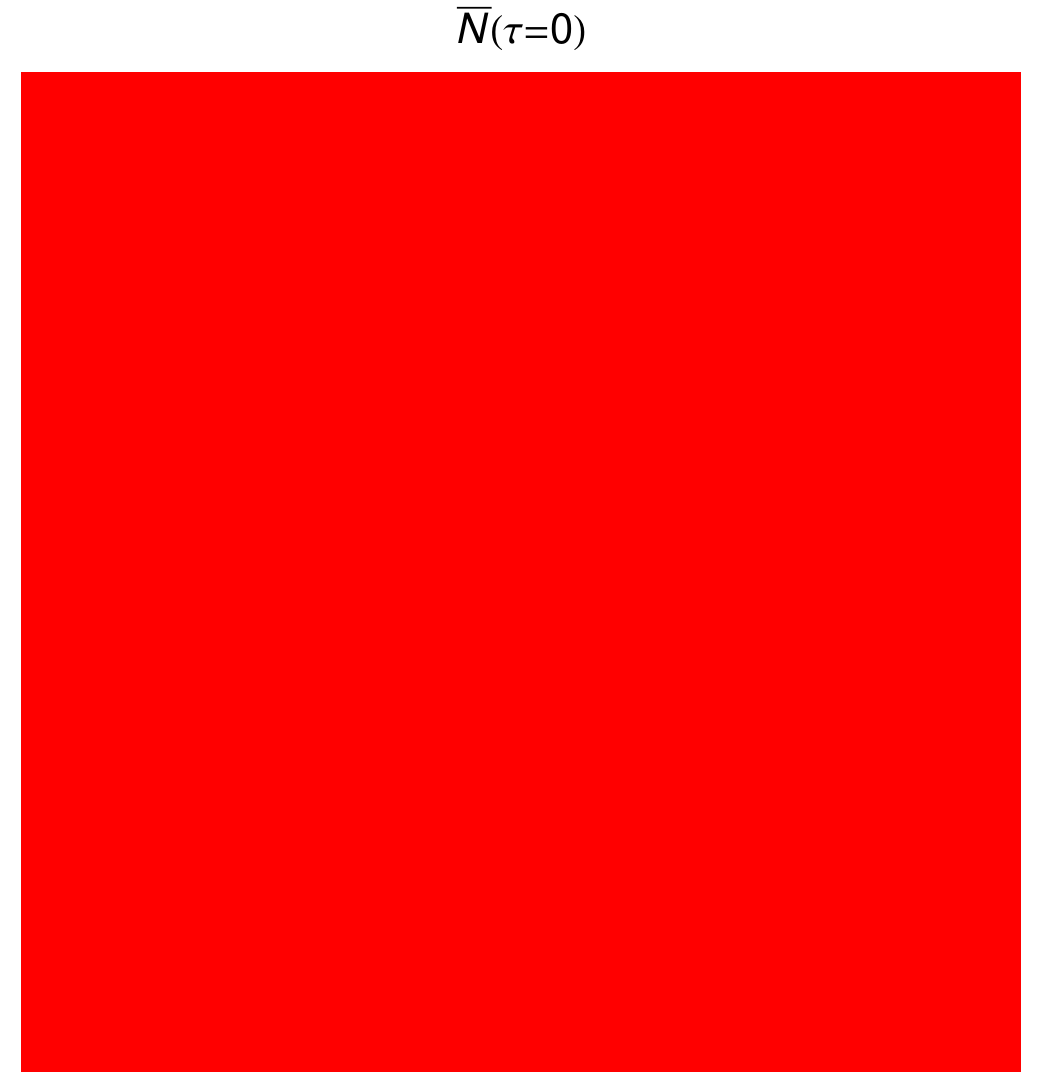}
\caption{Left panel shows, for $f=3$, the result of cooling a spacetime grid with random fluctuations around $\bar N\equiv (\bar N_2,\bar N_3)=(\frac 23,\frac 23)$ at large $\tau$ to $\tau=0$, where three phases  coexist: $\bar N = (0,0)$ (red squares), $\bar N=(\frac 12,1)$ (dark-blue squares), and $\bar N=(1,\frac 12)$ (light-green squares).
Center panel shows result starting from $\bar N=(\frac 34,\frac 34)$ at high $\tau$, which results in two coexisting phases ($\bar N=(\frac 12,1)$ and $\bar N=(1,\frac 12)$) at $\tau=0$.
Right panel shows results starting from $\bar N=(\frac 12,\frac 12)$ at high $\tau$, resulting in a single phase (characterized by $\bar N=(0,0)$) at $\tau=0$.}
\label{fig:f3-grid-inhom}
\end{figure}

All $\bar \mu$'s inside a given hexagonal cell result in identical $\bar N$ as $\tau\to 0$, given by the coordinates of the center of the cell. For example, $\bar \mu$'s in the central hexagonal cell lead to $\bar N_{2,3}=(0,0)$ at $\tau=0$, the six surrounding cells are characterized by $\bar N_{2,3}=\pm(1,\frac 12)$, $\bar N_{2,3}=\pm(\frac 12,1)$, and $\bar N_{2,3}=\pm(-\frac 12,\frac 12)$. 
Every vertex is common to three cells. The coordinates of the vertices between the central cell and the six surrounding cells are $\pm(\frac 23, \frac 23)$, $\pm(0,\frac 23)$, $\pm(\frac 23,0)$, $\pm(1,1)$, $\pm(0,1)$, $\pm(1,0)$. All other vertices in the $\bar\mu$ plane can be generated by shifts of the form \eqref{eq:f3-shift}. 

First-order phase transitions occur between neighboring cells with different particle numbers  $\bar N_{2,3}$ at $\tau=0$. At the edges of the hexagonal cells, two phases can coexist, and at the vertices, three phases can coexist at zero temperature. 

\begin{figure}[htb]
\centering
\includegraphics[width=0.7\textwidth]{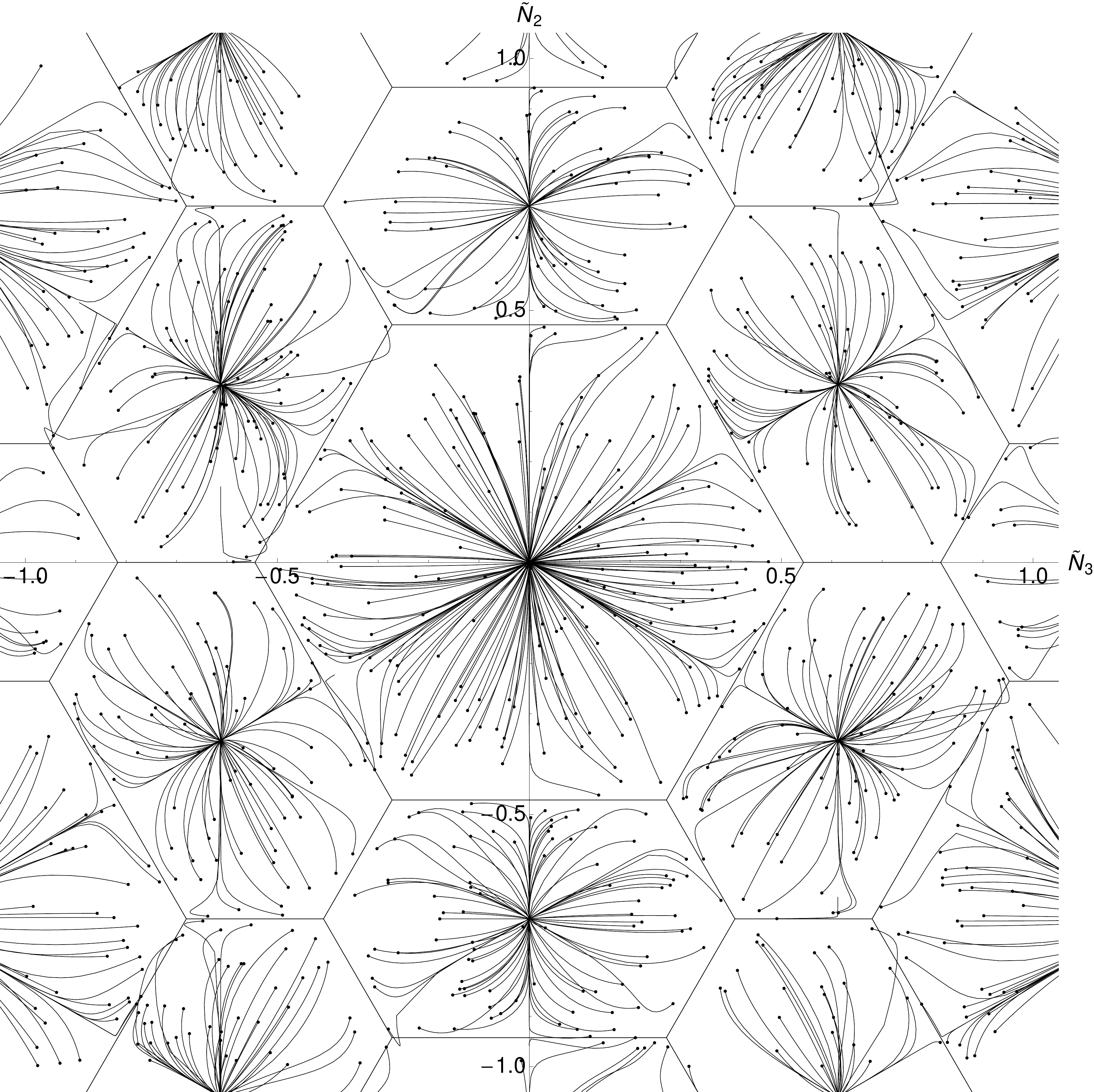}
\caption{Visualization of the $\tilde N$ evolution with decreasing $\tau$ starting from randomly scattered initial points at $\tau=\infty$ (indicated by dots in the plot).}
\label{fig:f3-flow}
\end{figure}

In analogy
to the two-flavor case (cf.~Fig.~\ref{fig:f2-grid-inhom}), a
high-temperature system with small fluctuations (as a function of
Euclidean spacetime) of $\bar \mu_{2,3}$ can result in two or three phases
coexisting  or result in a pure state as $\tau\to 0$ depending on the
choice of $\bar\mu_{2,3}$ (see
Fig.~\ref{fig:f3-grid-inhom} for examples of all three cases).
Figure~\ref{fig:f3-flow} shows the flow of $(\tilde N_3(\tau),\tilde N_2(\tau))$ from $\tau=\infty$ to $\tau=0$ at fixed $(\tilde \mu_3,\tilde \mu_2)=(\tilde N_3(\tau=\infty),\tilde N_2(\tau=\infty))$. The zero-temperature limit $(\tilde N_3(0),\tilde N_2(0))$  is given by the coordinates of the center of the respective hexagonal cell.

\subsection{f=4}

\begin{figure}[htb]
\centering
\includegraphics[width=0.3\textwidth]{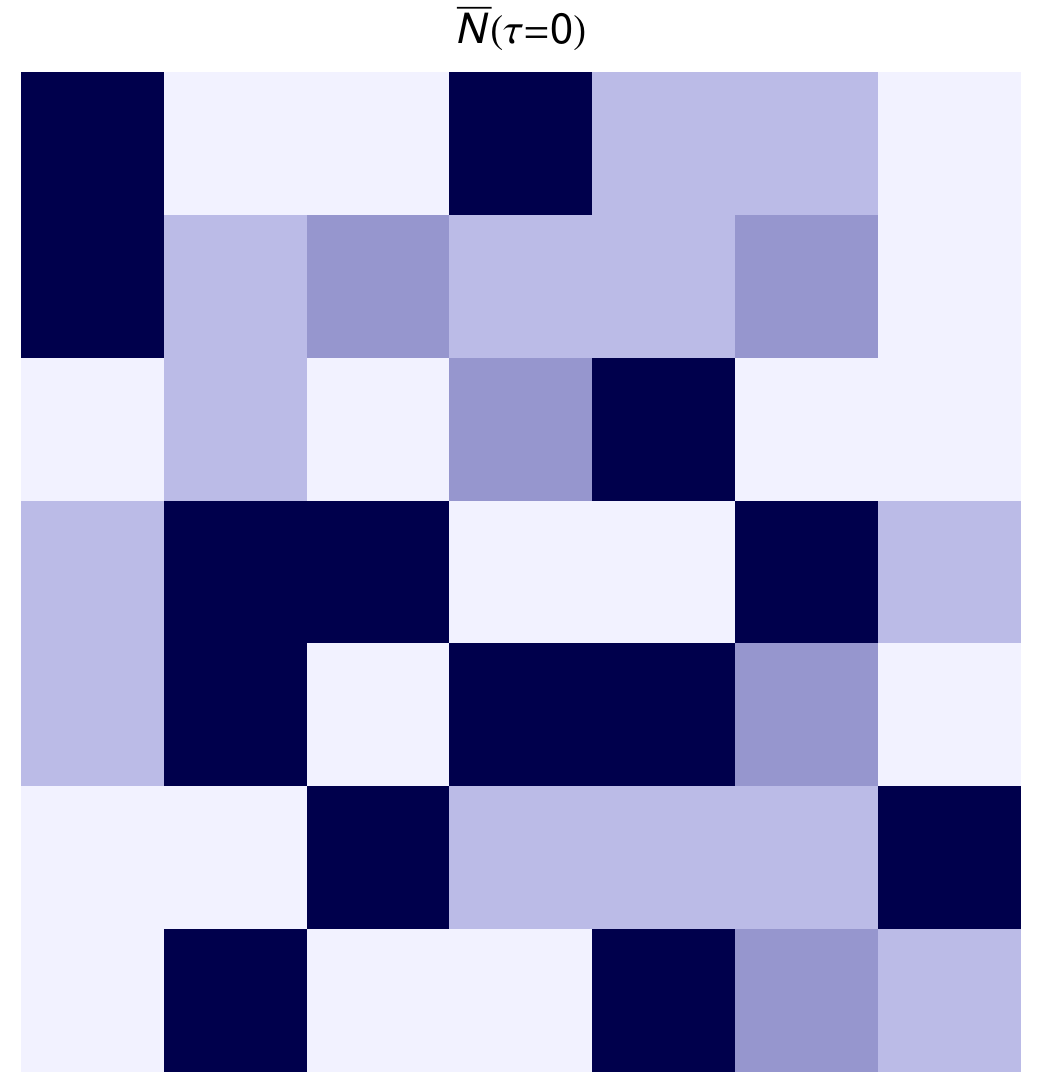}\qquad
\includegraphics[width=0.3\textwidth]{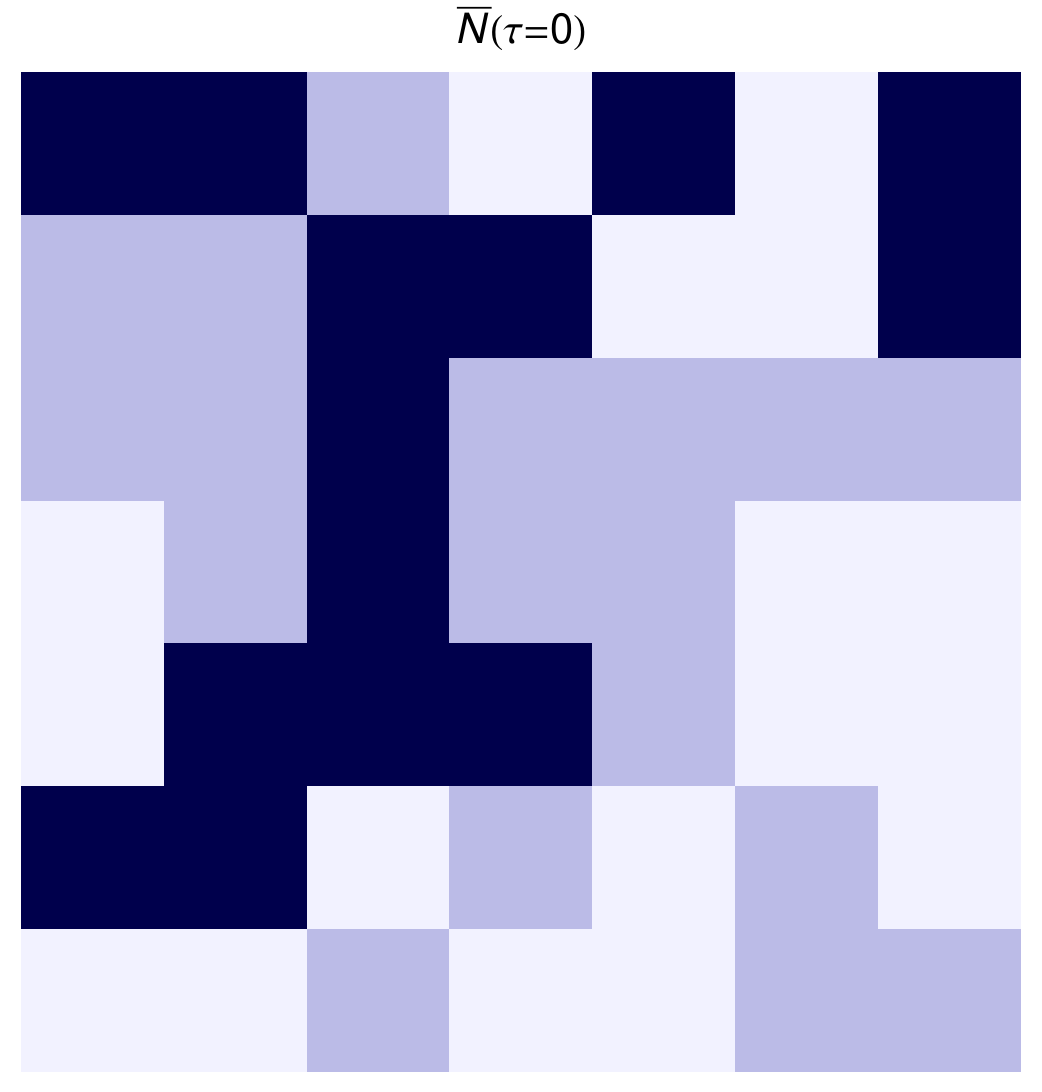}\qquad
\includegraphics[width=0.3\textwidth]{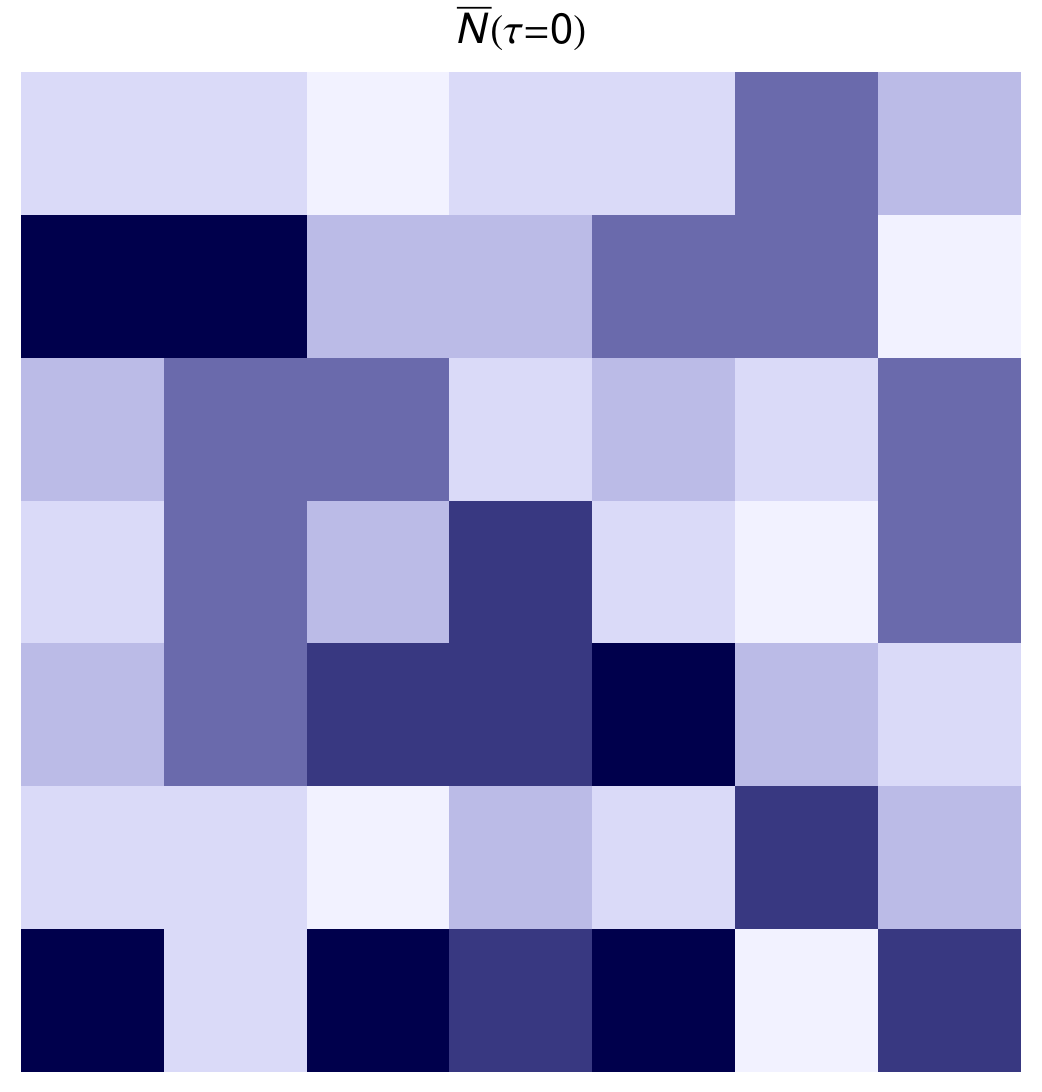}
\caption{Left panel shows, for $f=4$, the result of cooling a spacetime grid with random fluctuations around $\bar N\equiv (\bar N_2,\bar N_3, \bar N_4)=(\frac 34,\frac 34, \frac 34)$ at large $\tau$ to $\tau=0$, where four phases  coexist: $\bar N = (0,0,0)$, $\bar N=(\frac 12,\frac 12, 1)$, $\bar N=(\frac 12, 1, \frac 12)$, and $\bar N=(1, \frac 12, \frac 12)$. Different colors are assigned to different phases.
Center panel shows result starting from $\bar N=(\frac 78,\frac 78,\frac 78)$ at high $\tau$, which results in three coexisting phases ($\bar N=(\frac 12,\frac 12, 1)$, $\bar N=(\frac 12, 1, \frac 12)$, and $\bar N=(1, \frac 12, \frac 12)$) at $\tau=0$.
Right panel shows results starting from $\bar N=(1,1,1)$ at high $\tau$, resulting in six coexisting phases characterized by $\bar N=(\frac 12,\frac 12, 1)$, $\bar N=(\frac 12, 1, \frac 12)$, $\bar N=(1, \frac 12, \frac 12)$, $\bar N=(\frac 32,\frac 32, 1)$, $\bar N=(\frac 32, 1, \frac 32)$, and $\bar N=(1, \frac 32, \frac 32)$.}
\label{fig:f4-grid-inhom}
\end{figure}

We use Eq.~\eqref{zffinal} to identify the phase structure in the $(\bar \mu_2,\bar \mu_3, \bar \mu_4)$ space, which is divided into three-dimensional cells characterized by identical particle numbers $\bar N_{2,3,4}$ at zero temperature. At the boundaries of these cells, multiple phases can coexist at zero temperature (see Fig.~\ref{fig:f4-grid-inhom} for examples). We find different types of vertices (corners of the cells), where four and six phases can coexist. At all edges, three phases can coexist. 

We set $l_1=m_1+m_2-m_3$, $l_2=m_1$ and $l_3=m_2$.
From Eq.~\eqref{eq:shift-mu} for $f=4$, we see that the phase structure is periodic under
\begin{align}\label{eq:f4-shift}
\begin{pmatrix}
\bar \mu_2 \\
\bar \mu_3 \\
\bar \mu_4
\end{pmatrix} \rightarrow 
\begin{pmatrix}
\bar \mu_2 \\
\bar \mu_3 \\
\bar \mu_4
\end{pmatrix}
+
l_1
\begin{pmatrix}
1 \\
1 \\
0 
\end{pmatrix} 
+
l_2
\begin{pmatrix}
1 \\
0 \\
1 
\end{pmatrix} 
+
l_3
\begin{pmatrix}
0 \\
1 \\
1 
\end{pmatrix} 
\qquad\qquad l_{1,2,3}\in\Z\,.
\end{align}

\begin{figure}[htb]
\centering
\includegraphics[width=0.47\textwidth]{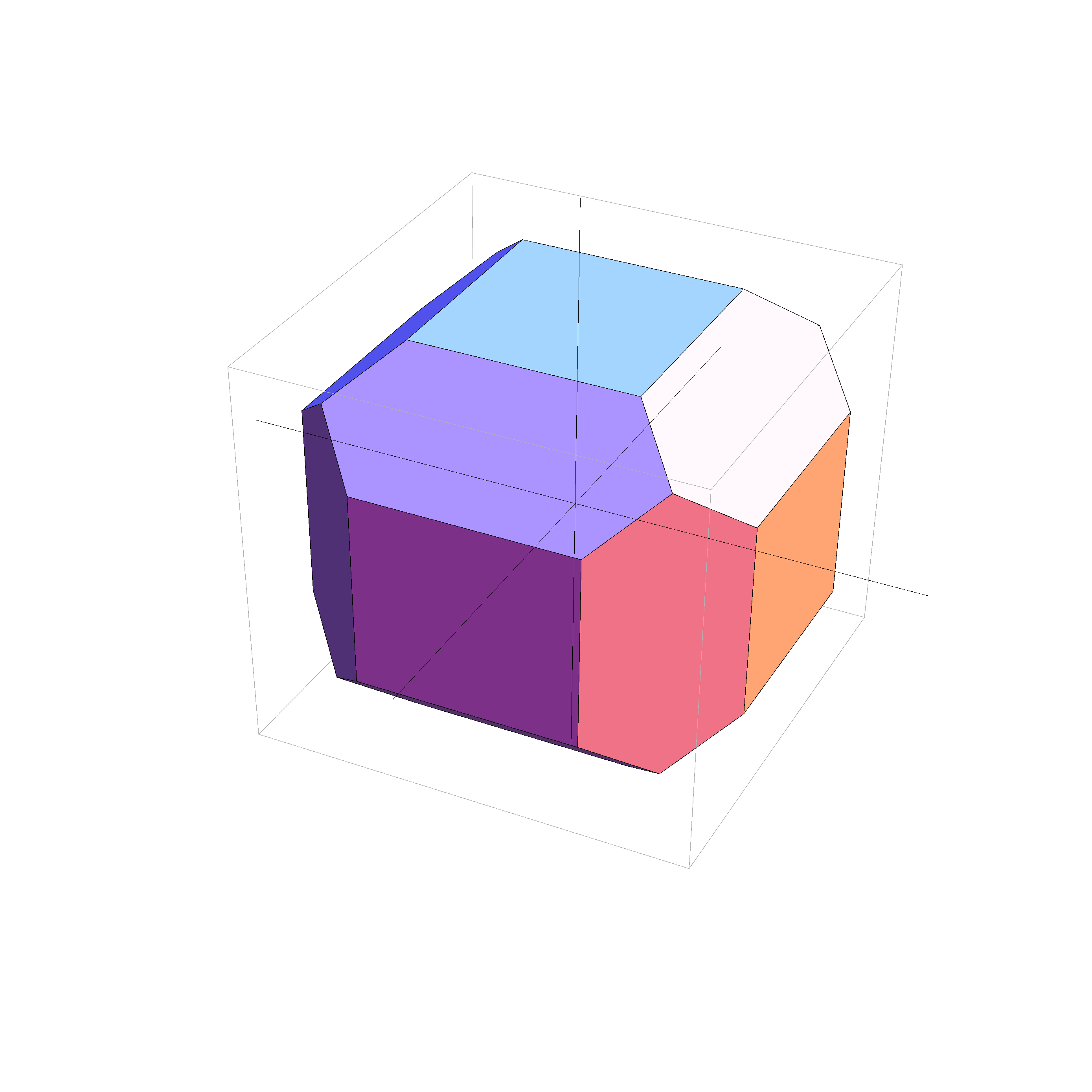}\hfill
\includegraphics[width=0.47\textwidth]{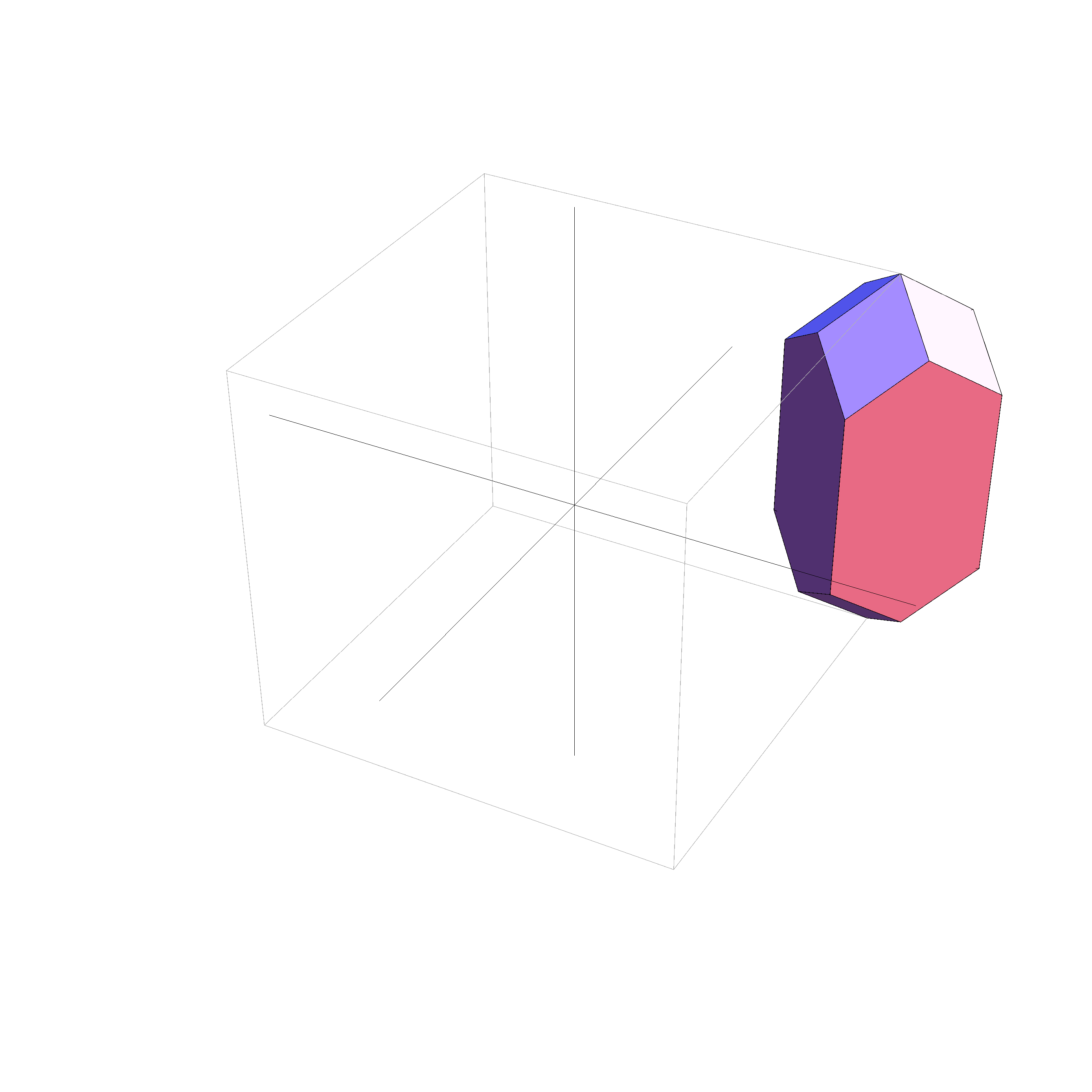}\\
\includegraphics[width=0.47\textwidth]{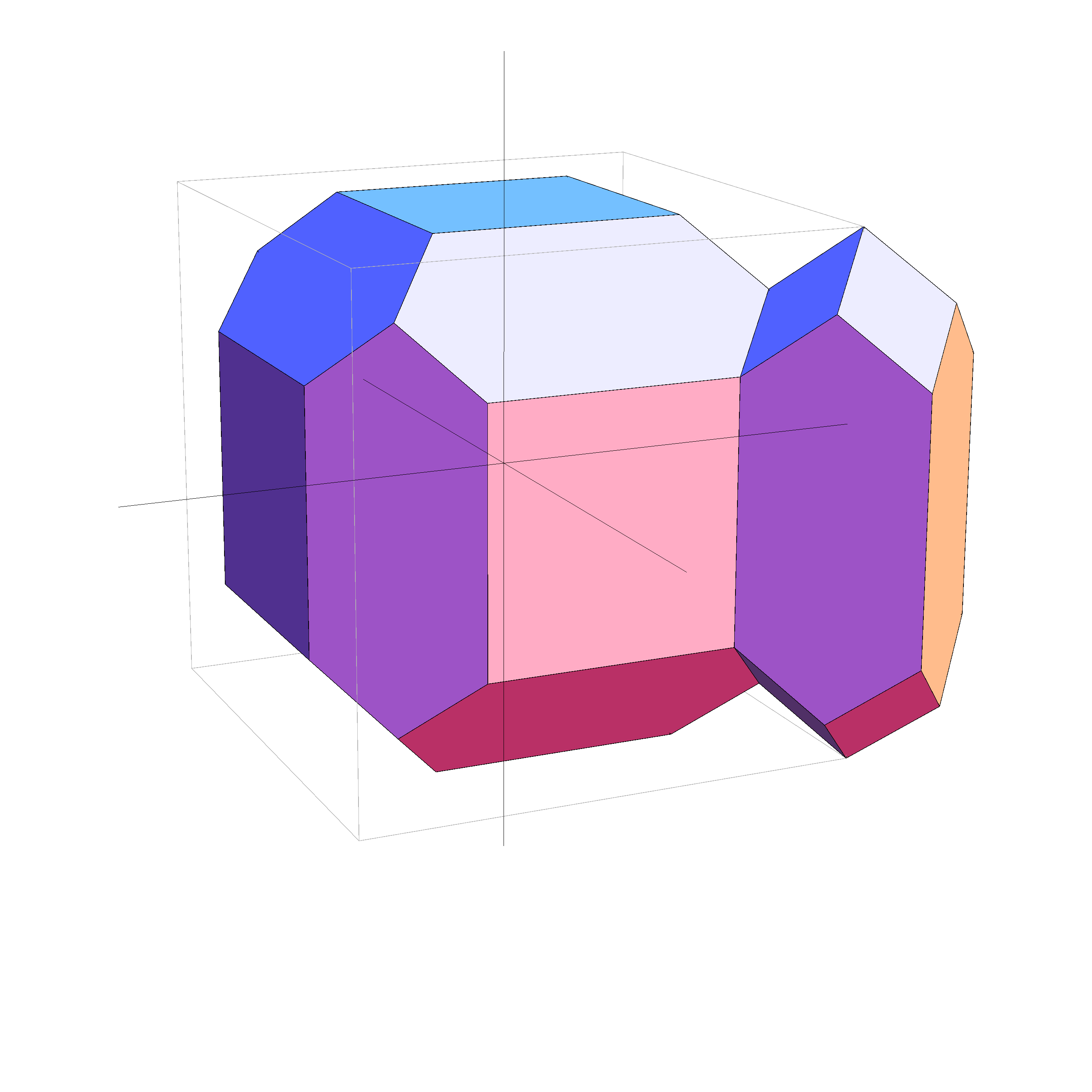}\hfill
\includegraphics[width=0.47\textwidth]{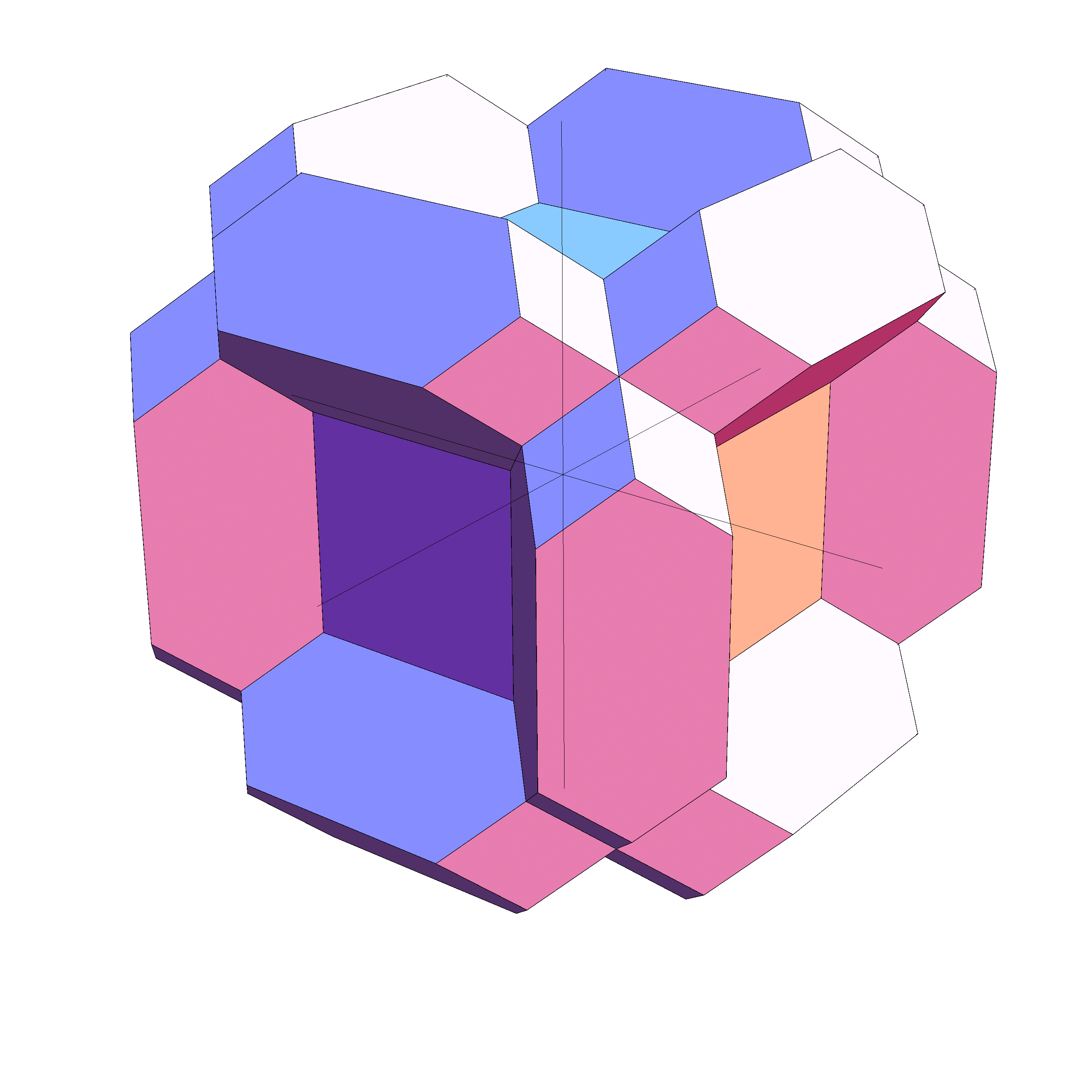}\\
\caption{Cells defining the zero-temperature phase structure for $f=4$ in the $\tilde \mu$ coordinates as described in the text. The top left figure shows the central ``cubic'' cell, the top right figure a single ``edge'' cell. The bottom right figure shows the cubic cell together with all 12 attaching edge cells.}
\label{fig:f4-cells}
\end{figure}

As in the three flavor case, we observe that the phase structure exhibits higher symmetry in coordinates $\tilde \mu$ which are related to $\mu$ through an orthonormal transformation. A particularly convenient choice for $f=4$ turns out to be given by
\begin{align}\label{eq:trafo-f4}
\begin{pmatrix}
\tilde \mu_1 \\
\tilde \mu_2 \\
\tilde \mu_3 \\
\tilde \mu_4
\end{pmatrix} = 
\frac 12 
\begin{pmatrix} 1 & 1 \\ 1 & -1 \end{pmatrix} \otimes 
\begin{pmatrix} 1 & 1 \\ 1 & -1 \end{pmatrix} 
\begin{pmatrix}
 \mu_1 \\
 \mu_2 \\
 \mu_3 \\
 \mu_4
\end{pmatrix}\,,
\end{align}
since the phase structure becomes periodic under shifts parallel to the coordinate axes: 
\begin{align}\label{eq:f4-shift-tilde}
\begin{pmatrix}
\tilde \mu_2 \\
\tilde \mu_3 \\
\tilde \mu_4
\end{pmatrix} \rightarrow 
\begin{pmatrix}
\tilde \mu_2 \\
\tilde \mu_3 \\
\tilde \mu_4
\end{pmatrix}
+
l_1
\begin{pmatrix}
0 \\
0 \\
1 
\end{pmatrix} 
+
l_2
\begin{pmatrix}
1 \\
0 \\
0 
\end{pmatrix} 
+
l_3
\begin{pmatrix}
0 \\
1 \\
0 
\end{pmatrix} 
\qquad\qquad l_{1,2,3}\in\Z
\end{align}
as obtained from Eq.~\eqref{eq:f4-shift}. 
An alternative representation of the partition function in these coordinates is given in Eq.~\eqref{eq:Zt-alt-f4}.
At zero temperature the $\tilde \mu_{2,3,4}$ space is divided into two types of cells which are characterized by identical particle numbers  (see Fig.~\ref{fig:f4-cells} for visualizations). We can think of the first type as a cube (centered at the origin, with side lengths 1 and parallel to the coordinate axes) where all the edges have been cut off symmetrically. The original faces are reduced to smaller squares (perpendicular to the coordinate axes) with corners at $\tilde \mu_{2,3,4}=(\pm \frac 12, \pm \frac 14, \pm \frac 14)$ (permutations and sign choices generate the six faces). This determines the coordinates of the remaining 8 corners to be located at $(\pm \frac 38, \pm \frac 38, \pm \frac 38)$. The shift symmetry \eqref{eq:f4-shift-tilde} tells us that these ``cubic'' cells are stacked together face to face. 
The remaining space (around the edges of the original cube) is filled by cells of the second type (in the following referred to as ``edge'' cells), which are identical in shape and are oriented parallel to the three coordinate axes.

This leads to different kinds of vertices (at the corners of the cells described above) where multiple phases can coexist at zero temperature. There are corners which are common points of two cubic and two edge cells (coexistence of 4 phases, for example at $(\pm \frac 12, \pm \frac 14, \pm \frac 14)$), there are corners which are common points of one cubic and three edge cells (coexistence of 4 phases, for example at $(\pm \frac 38, \pm \frac 38, \pm \frac 38)$), and there are corners which are common points of six edge cells (coexistence of six phases, for example at $\tilde \mu_{2,3,4}=(\pm \frac 12,\pm \frac 12,\pm \frac 12)$. Any edge between two of these vertices is common to three cells.

\subsection{\boldmath Phase structure for $f>4$}

For $f=3$ and $f=4$, we find that the coordinates $(\bar \mu_2,\ldots,\bar \mu_f)$ of all vertices (corners of the cells in the $\bar \mu$ space resulting in identical particle numbers  at zero temperature) are multiples of $\frac 1 f$. 
In general, two special vertices are located at $\bar \mu_i=1$ for all $2\leq i\leq f$ and $\bar \mu_i=1-\frac 1f$ for all $2\leq i\leq f$.

If $\bar \mu_i=1-\frac 1f$ for all $2\leq i \leq f$, we find that $f$ phases can coexist at zero temperature. These have particle numbers $\bar N_{2,\ldots,f}=(0,\ldots,0)$ and all $(f-1)$ distinct permutations of $(1,\frac 12,\ldots ,\frac 12)$.
 
If $\bar \mu_i=1$ for all $2\leq i \leq f$, we find that $f \choose 2$ phases can coexist at zero temperature. The corresponding particle numbers are given by the $(f-1)$ distinct permutations of $(1,\frac 12,\ldots ,\frac 12)$ and the $f-1 \choose 2$ distinct permutations of $(\frac 32,\frac 32,1,\ldots,1)$.

While for $f=5$, we find only up to $5 \choose 2$ coexisting phases,
we find up to $6 \choose 3$ coexisting phases for $f=6$ (for example
at $\bar \mu_{2,\ldots, 6}=(1,\frac{1}{2},0,0,0)$). We also find up to
$8\choose 4$ coexisting phases for $f=8$ (for example at $\bar \mu_{2,\ldots, 8}=(1,1,1,1,1,1,0)$).
This leads us to conjecture that the maximal number of coexisting phases is given by $f \choose{ \lfloor f/2 \rfloor} $, increasing exponentially for large $f$.

\section{Conclusions}
Multiflavor QED in two dimensions with flavor-dependent chemical potentials
exhibits a rich phase structure at zero temperature. We studied 
massless multiflavor QED on a two-dimensional tours. The system is
always in a state with a net charge of zero in the Euclidean
formalism due to the integration over the toron variables. The toron
variables
completely dominate the dependence on the chemical potentials and the
resulting
partition function has a representation in the form of a
multidimensional theta function. We explicitly worked out the two-dimensional phase
structure for the three flavor case and the three-dimensional phase
structure
for the four flavor case. The different phases at zero temperature are characterized by certain values of the particle numbers  and separated by first-order phase transitions. We showed that two or three phases can coexist in
the case of three flavors. We also showed that two, three, four and
six phases can coexist in the case of four flavors. Based on our
exhaustive studies of the three and four flavor case and an
exploratory
investigation of the five, six, and eight flavor case we conjecture
that up to $f \choose{ \lfloor f/2 \rfloor} $ phases can coexist in a
theory with $f$ flavors.

\appendix

\section{Alternative representations of the partition function}\label{sec:Zalt}

There are many equivalent representations of the partition function $Z_t(\bmu,\tau)$, related by variable changes of the integer summation variables in \eqref{toronpf}, \eqref{maineqn} or \eqref{zffinal}. Here we present the result obtained by an orthonormal variable change at the level of Eq.~\eqref{toronpf}, splitting the chemical potentials $\mu_1,\ldots,\mu_f$ in one flavor-independent and $(f-1)$ traceless components according to
\begin{align}
\tilde \mu_1&=\frac 1{\sqrt f}\sum_{i=1}^f \mu_i\,,\label{eq:orthonormal1}\\
\tilde \mu_j&=\frac 1{\sqrt{j(j-1)}}\left(\sum_{i=1}^{j-1} \mu_i -(j-1) \mu_j\right)\,, \qquad 2\leq j\leq f \,.\label{eq:orthonormal2}
\end{align}
The induced variable change in the $2f$ integer summation variables in Eq.~\eqref{toronpf} is non-trivial and requires successive transformations of the form
\begin{align}\label{eq:int-trafo}
\sum_{k,l=-\infty}^\infty f(k-l,M k+l)=\sum_{q=0}^{M} \sum_{m,n=-\infty}^\infty f((M+1) m +q, (M+1) n -q)\,,\qquad M\in\N_+\,.
\end{align}
In this way, it is possible to write the partition function as a product of $2f-2$ one-dimensional theta functions, where $f-1$ factors are independent of the chemical potentials and each one of the other $f-1$ factors depends only on a single traceless chemical potential $\tilde \mu_i$ (with $2\leq i\leq f$). However, the arguments of the theta functions are not independent since they involve a number of finite summation variables resulting from variable changes of the form \eqref{eq:int-trafo} and the partition function does not factorize. The final result reads
\begin{align}\label{eq:Zt-orthonormal}
Z_t(\bmu,\tau)&\propto \left(\prod_{j=1}^f \sum_{k_j=0}^1\right) \left(\prod_{j=2}^f \sum_{q_j=0}^{j-1} \sum_{p_j=0}^{j-1}\right)\delta_{0,\left(2 \sum_{j=2}^f p_j + \sum_{j=1}^f k_j \right)\,\text{mod}\, 2f}\cr
&\quad\times 
\left[\prod_{j=2}^f 
h_{2 \tau j(j-1)}\left(\frac1{j(j-1)}\sum_{i=2}^{j-1} q_i -\frac 1j q_j+\frac1{\sqrt{j(j-1)}}\left(\frac{\tilde k_j}2-\frac i\tau \tilde\mu_j\right) \right)
\right]\cr
&\quad\times 
\left[\prod_{j=2}^f 
h_{2 \tau j(j-1)}\left(\frac1{j(j-1)}\sum_{i=2}^{j-1} p_i -\frac 1j p_j+\frac1{\sqrt{j(j-1)}} \frac{\tilde k_j}2 \right)
\right]\,,
\end{align}
where $\tilde k_j=\frac 1{\sqrt{j(j-1)}}\left(\sum_{i=1}^{j-1} k_i -(j-1) k_j\right)$ and 
\begin{align}
h_\alpha(z)\equiv \sum_{n=-\infty}^\infty e^{-\pi \alpha (n+z)^2}\,.
\end{align}
Permuting indices in variable changes of the form \eqref{eq:orthonormal2} shows that the $(f-1)$-dimensional finite sum $\prod_j \sum_{p_j}$ will result in an expression that depends only on $\sum_{j=1}^f k_j$.  

To study the zero-temperature properties, we can apply the Poisson summation formula for each factor of $h_\alpha(z)$ in Eq.~\eqref{eq:Zt-orthonormal}.

\subsection{\boldmath Explicit form for $f=3$}

For $f=3$, the Poisson-resummed version of \eqref{eq:Zt-orthonormal} can be simplified to
\begin{align}
Z_t(\bmu,\tau) &\propto \sum_{m_1,m_2,l_1,l_2=-\infty}^\infty \delta_{0,\left(m_1+l_2\right)\,\text{mod}\,2}\, \delta_{0,\left(m_2+l_1\right)\,\text{mod}\,2}\,\,
e^{-\frac{\pi}{4\tau}\left( (m_1+m_2)^2 +3 (m_1-m_2)^2 +(l_1+l_2)^2 +\frac 13 (l_1-l_2)^2\right)}\cr
&\quad\times e^{\frac{\pi}\tau \left( (m_1+m_2)\sqrt{2} \tilde\mu_2 + (m_1-m_2) \sqrt{6} \tilde \mu_3 \right) }\,.
\end{align}
For $\tau \to 0$, the sums over $l_{1,2}$ become trivial and we obtain
\begin{align}\label{eq:Zt-alt-f3}
Z_t(\bmu,\tau) \rightarrow \sum_{m_1,m_2=-\infty}^\infty e^{ -\frac \pi \tau\left(
\frac 14 (m_1+m_2)^2 +\frac 34 (m_1-m_2)^2 -(m_1+m_2) \sqrt{2} \tilde \mu_2 + (m_1-m_2)\sqrt{6} \tilde \mu_3
+\frac 13 \left(1-\delta_{0,m_1\text{mod} 2} \delta _{0, m_2 \text{mod} 2}\right) 
\right)}\,.
\end{align}
The particle numbers  $\tilde N_2$ and $\tilde N_3$ at zero temperature are determined by those integer pairs $(m_1,m_2)$ dominating the sum in Eq.~\eqref{eq:Zt-alt-f3}. Compared to the general expression in Eq.~\eqref{zffinal}, we have reduced the number of summation variables from four to two, which simplifies the search for vertices where multiple phases coexist. Furthermore, there is a one-to-one map from $(\tilde N_2,\tilde N_3)$ to $(m_1,m_2)$ inside any given cell in the zero-temperature phase-structure. Once we have located neighboring cells in terms of $(m_1,m_2)$, we can immediately read off the $\tilde \mu_{2,3}$ coordinates of the corresponding vertices/edges between them (by requiring that the contributions to the sum \eqref{eq:Zt-alt-f3} are identical).

\subsection{\boldmath Explicit form for $f=4$}

Following the general procedure described above, we can write the partition function for $f=4$ in the coordinates defined in Eq.~\eqref{eq:trafo-f4} as
\begin{align}\label{eq:Zt-alt-f4}
Z_t(\bmu,\tau)\propto \sum_{m_2,m_3,m_4,n_2,n_3,n_4=-\infty}^\infty \delta_{0,(m_2+m_3+m_4)\,\text{mod}\,2} \, \delta_{0,(m_2+n_2+n_3)\,\text{mod}\,2} \, \delta_{0,(m_3+n_3+n_4)\,\text{mod}\,2} \,\, e^{-\frac \pi{2\tau} \sum_{j=2}^4\left(m_j^2+n_j^2-4 m_j \tilde \mu_j\right)}\,.
\end{align}
Similarly to the three flavor case, the sum over $n_{2,3,4}$ becomes trivial in the $\tau\to0$ limit, depending only on $m_2\,\text{mod}\,2$ and $m_3\,\text{mod}\,2$. The remaining summation variables $m_{2,3,4}$ directly determine the particle numbers  in the different phases at zero temperature and the vertices can be found analogously to the three flavor case.

\begin{acknowledgments} 
The authors acknowledge partial support by the NSF under grant numbers 
PHY-0854744 and PHY-1205396. RL would like to acknowledge the theory group at BNL for
pointing out that we are computing traceless particle numbers and not traceless
number densities.
\end{acknowledgments}

\end{document}